\documentclass[10pt,twocolumn,letterpaper]{article}

\usepackage{iccv}
\usepackage{times}
\usepackage{epsfig}
\usepackage{graphicx}
\usepackage{amsmath}
\usepackage{amssymb}
\usepackage{subcaption}
\usepackage{comment}
\usepackage{multicol}
\usepackage{multirow}

\usepackage[breaklinks=true,bookmarks=false]{hyperref}

 \iccvfinalcopy 

\def\iccvPaperID{3899} 

\ificcvfinal\fi

\begin{document}

\title{Adversarial Mutual Leakage Network for Cell Image Segmentation}

\author{Hiroki Tsuda \qquad Kazuhiro Hotta\\
Meijo University\\
1-501 Shiogamaguchi, Tempaku-ku, Nagoya 468-8502, Japan\\
{\tt\small 193427019@ccalumni.meijo-u.ac.jp, kazuhotta@meijo-u.ac.jp}
}

\maketitle
\ificcvfinal\thispagestyle{empty}\fi

\begin{abstract}
We propose three segmentation methods using GAN and information leakage between generator and discriminator. First, we propose an Adversarial Training Attention Module (ATA-Module) that uses an attention mechanism from the discriminator to the generator to enhance and leak important information in the discriminator. ATA-Module transmits important information to the generator from the discriminator. Second, we propose a Top-Down Pixel-wise Difficulty Attention Module (Top-Down PDA-Module) that leaks an attention map based on pixel-wise difficulty in the generator to the discriminator. The generator trains to focus on pixel-wise difficulty, and the discriminator uses the difficulty information leaked from the generator for classification. Finally, we propose an Adversarial Mutual Leakage Network (AML-Net) that mutually leaks the information each other between the generator and the discriminator. By using the information of the other network, it is able to train more efficiently than ordinary segmentation models. Three proposed methods have been evaluated on two datasets for cell image segmentation. The experimental results show that the segmentation accuracy of AML-Net was much improved in comparison with conventional methods.

\end{abstract}

\section{Introduction}
\label{sec:introduction}

Automated cell image segmentation has been widely studied due to the large number of cell images and the tedious task of obtaining dense annotations.
Overall time and cost savings are expected to be achieved by automated processing without human involvement. Manual segmentation is slow and burdensome to process, and thus there is a significant demand for algorithms that can perform segmentation quickly and accurately without human intervention. However, cell image segmentation is a difficult task because the number of supervised images is smaller and there is not regularity compared to the other datasets such as automatic driving. A large number of supervised images require manual labeling which take a lot of effort and time. Therefore, it is necessary to improve the segmentation ability for pixel-level recognition with small number of training images.

By the advent of Convolutional Neural Networks (CNNs), near-human level performance can be achieved in medical image analysis tasks such as blood vessel extraction from fundus images~\cite{hoover2000STARE,owen2009CHASE,2004DRIVE}, cancerous lung nodule detection~\cite{liao2019evaluate}, and cell image segmentation~\cite{imanishi2018novel,zheng2018wbc}. CNNs have become the de facto standard in the field of image recognition, Fully Convolutional Networks~(FCNs) and U-Net~\cite{unet} are the commonly used for segmentation.

Generative Adversarial Network (GAN)~\cite{gan} is for image generation. It preforms adversarial training between generator and discriminator. The discriminator enchants the difference between real and generated images. The adversarial training allows to generate realistic images. In addition, pix2pix~\cite{pix2pix} is the extended version of GAN can train image-to-image translation. In particular, pix2pix is also effective for semantic segmentation tasks~\cite{luc2016semantic}.

\begin{figure}[t] 
    \centering
    \includegraphics[width=0.95\linewidth]{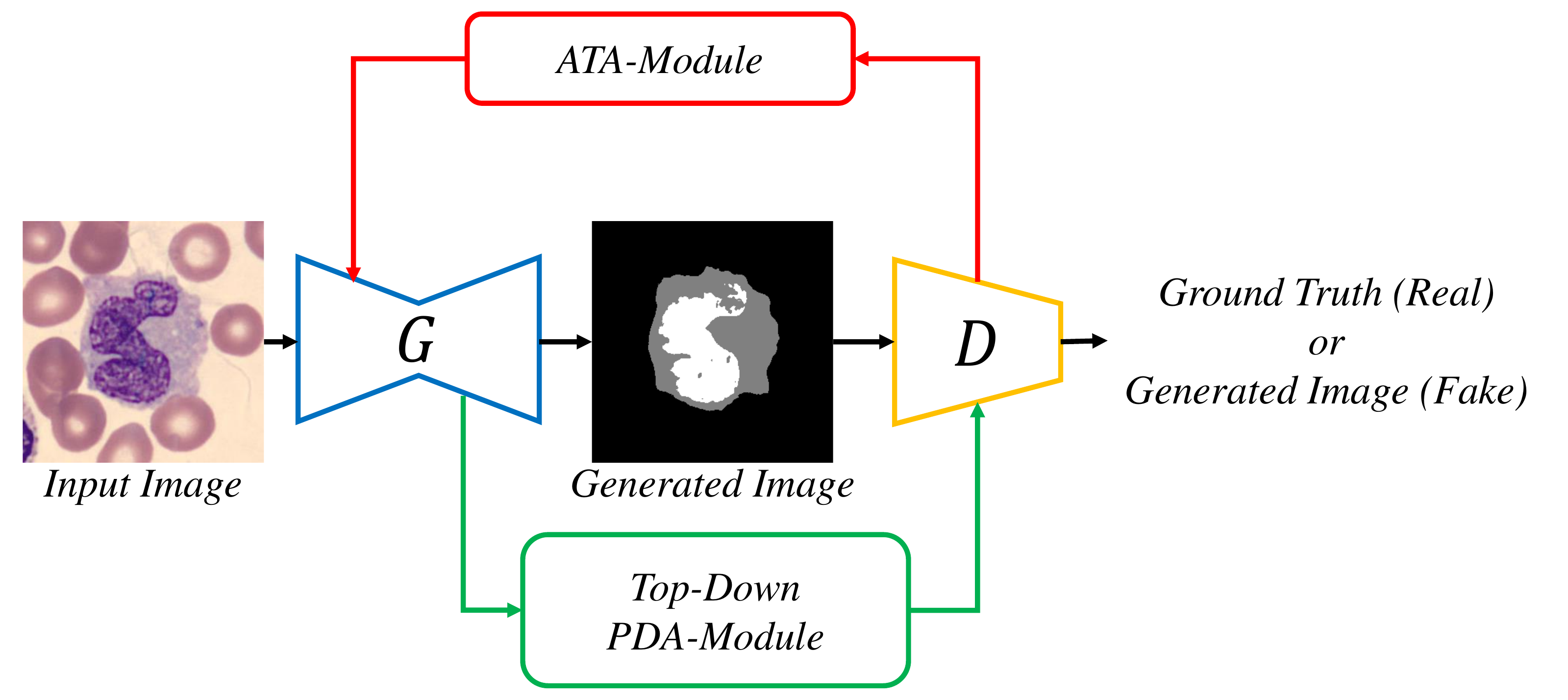}
  \caption{\textbf{Overview of Adversarial Mutual Network.} Adversarial training is composed of two components; the generator ($G$) and the discriminator ($D$). We introduce two attention mechanisms that leak information into adversarial training, and train them using each other's information.}
  \label{fig:overall}
\end{figure}

We focused on the relationship between the generator and the discriminator. We consider that adversarial training could be more efficient by mutually leaking important information from the generator and the discriminator. In this paper, we propose
Adversarial Mutual Leakage Network (AML-Net) which consists of two new attention modules. 
The first one is Adversarial Training Attention Module (ATA-Module) that creates an attention map from the feature map in discriminator and leaks it to the generator.
The second one is Top-Down Pixel-wise Difficulty Attention (Top-Down PDA) Module that creates an attention map based on pixel-wise difficulty of generator and leaks it to the discriminator. AML-Net combines these two attention mechanisms and leaks them to each other. Figure~\ref{fig:overall} shows the overview of AML-Net, which aims to improve the performance of generator and discriminator by mutually leaking information through two attention mechanisms. 

In experiments on two kinds of cell image datasets ~\cite{sstem,zheng2018wbc}, we evaluate the proposed method. We confirmed that our method gave higher accuracy than conventional methods without information leakage. 

\section{Related works}
\label{sec:RelatedWorks}
\subsection{Semantic Segmentation}
FCNs based methods have achieved significant results for semantic segmentation. The original FCN~\cite{fcn} finally created high-dimensional feature map with low-resolution. This feature map has semantic information but fine information such as small objects and correct location are lost. Thus, if upsampling is used at the final layer, the accuracy is not sufficient. 
SegNet~\cite{segnet}, U-Net~\cite{unet} and RefineNet~\cite{lin2017refinenet} obtained feature map with high-resolution. They are designed to gradually recover the spatial information by decoder to combine various contextual information extracted by encoder. In addition, Attention U-Net~\cite{oktay2018attention_unet} introduces attention to U-Net. By adding a per-pixel attention gate similar to the sSE block~\cite{roy2018scsenet} to the skip connection, it improves the segmentation accuracy. We improved the segmentation accuracy by leaking the information through attention map between the generator and the discriminator.

Alternatively, DeepLabv3+~\cite{deeplabv3plus} removes the last two downsampling processes from ResNet~\cite{resnet} and introduces dilated (atrous) convolutions~\cite{dilatedconv} to maintain the receptive field. As a result, it can hold location information with sufficient size. Furthermore, DeepLabv3+~\cite{deeplabv3plus} combined the advantages of the encoder-decoder structure of DeepLabv3~\cite{deeplabv3} to reduce the computational cost. In addition, FastFCN~\cite{wu2019fastfcn} used the original ResNet in encoder and adopted Joint Pyramid Upsampling to reduce the computational cost without degrading the performance.

Increasing the receptive field without decreasing the resolution by dilated convolution is that the resolution of the feature map is relatively large. This increases the computational cost, and the processing time for training and inference is long. Therefore, we studied the method for training important features based on U-Net, which is easy to use with simple computational resources and is widely used in medical and biological imaging.

\subsection{Adversarial Training}

GANs~\cite{gan} has achieved success in image generation tasks, including image-to-image translation~\cite{pix2pix,liu2016coupled,park2019semantic,taigman2017unsupervised,xue2018segan,cycle}, domain adaptation~\cite{choi2019self,hoffman2017cycada,saito2018maximum,tzeng2017adversarial} and text-to-image synthesis~\cite{reed2016generative,zhang2017stackgan}. Adversarial training improved image generation by training generator and discriminator to compete with each other. The discriminator trains to classify whether an input is a real or a generated image. On the other hand, the generator trains so that the generated image and the real image are not classified by the discriminator. Adversarial training network gave superior result on segmentation in comparison with non-adversarial deep networks~\cite{luc2016semantic}

Sawada et al.~\cite{sawada2018} has proposed a method to train more efficiently using the framework of adversarial training. By sending the feature maps in discriminator to generator, the segmentation accuracy of generator was improved. They concatenated the feature maps in discriminator to generator because the feature maps in discriminator include the difference between generated results and ground truth. However, the training parameters of the method are very large due to the usage of multiple generators and discriminators. In addition, since they did not select
the information from discriminator, the effect of leakage is not fully exploited.
We extract only the important information from the feature maps in generator and discriminator, and leak them each other between the generator and discriminator.

\subsection{Attention Mechanism}

Attention mechanism is
used in computer vision and natural language processing. In image recognition, important parts or channels are enhanced.
Residual Attention Network~\cite{wang2017residual} introduced a stacked network structure composed of multiple attention components, and attention residual training used residual training~\cite{resnet} in attention mechanism.
Squeeze-and-Excitation Network (SENet)~\cite{senet} introduced an attention mechanism that enhances the channels of feature maps. Accuracy booster blocks~\cite{accuracy-booster} and efficient channel attention module~\cite{wang2019eca} made further improvements by changing the fully-connected layer in SENet~\cite{senet}.
Transformer~\cite{transformer} performed the translation task only with the attention mechanism in the natural language processing. There are Self-Attention that uses the same tensor, and Source-Target-Attention that uses two different tensors. 
Several networks have been proposed that use Self-Attention to train the similarity between pixels in feature maps~\cite{fu2019dual,huang2019ccnet,stand-alone,wang2018non_local,sagan}. 
Pixel-wise Difficulty Attention (PDA) Module~\cite{matsuzuki2019PDA} is an attention mechanism that uses the confidence at each pixel as difficulty level. It trains efficiently by reinforcing pixels with low confidence as high difficulty. However, PDA-Module has a problem that the difficulty level becomes low when mis-classification is occurs with high confidence. 


In this paper, we have designed two kinds of attention mechanisms for better information leakage. The discriminator-to-generator attention mechanism is designed based on self-attention, and we aim to select effective information to the generator from the discriminator. The generator-to-discriminator attention mechanism creates and leaks an attention map with the probability of the correct class as the difficulty level by referring to ground truth. This attention mechanism is used in only training. Since pixel-wise difficulty becomes small through training, we do not need to use it in test phase.

\section{Proposed Method}

We present two attention mechanisms that leak information; the first mechanism leaks the information from the discriminator to the generator, and the second one leaks the information from the generator to the discriminator.
Finally, we combines the two attention mechanisms based on information leakage between the generator and discriminator.

\begin{figure}[t] 
    \centering
    \includegraphics[width=0.95\linewidth]{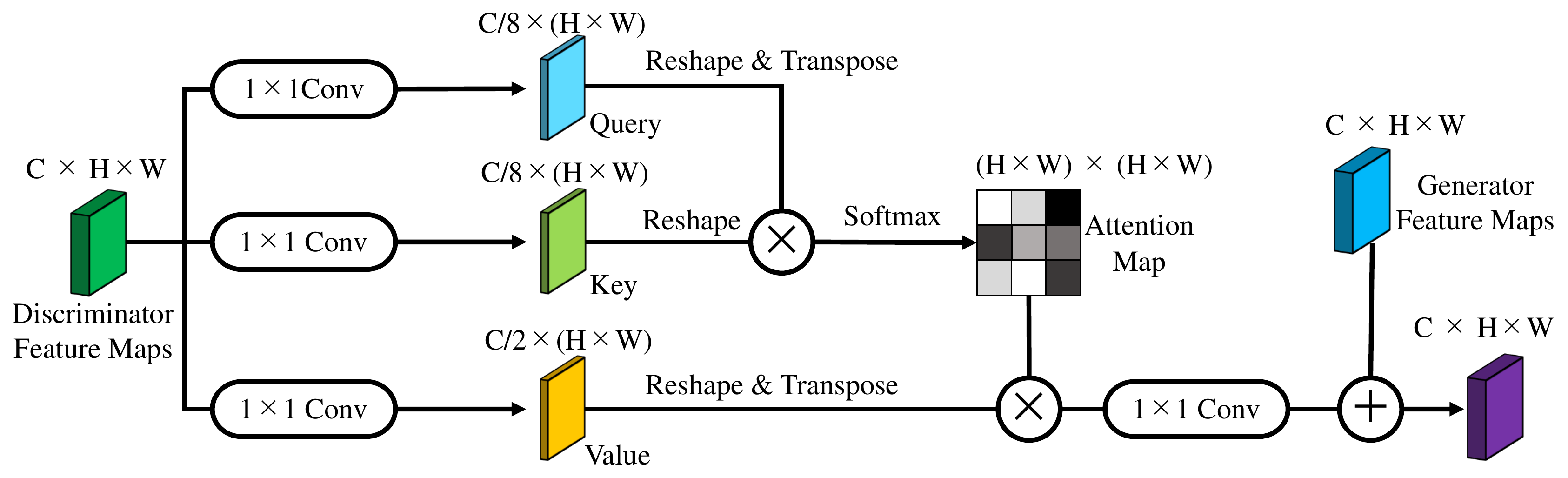}
  \caption{\textbf{Network structure of Adversarial Training Attention Module.} An attention map created from the discriminator's feature maps is leaked to the generator.}
  \label{fig:ATA-Module}
\end{figure}

\subsection{Adversarial Training Attention Module}

Conventional method leaks from the first discriminator to the second generator to improve accuracy~\cite{sawada2018}, but it uses multiple generators, 
and the feature map in the discriminator is just concatenated to the feature map in the the second generator. The good information for improving the generator is not extracted. 
In addition, the number of parameters increases in the method due to more than two generators. In order to increase the effect of leakage from the discriminator to the generator, we enhance the important pixels from the discriminator's feature map and then transfer them to the generator for efficient training. We propose an Adversarial Training Attention Module (ATA-Module) that enhances the important pixels of the feature map in the discriminator and then leaks them to the generator.

The discriminator determines whether the generated image is real or fake. The feature map in the discriminator has the information about the difference between the generated image and ground truth. 
The feature map in the discriminator contains the information on what is wrong with the generated image when it is judged as fake.
The segmentation accuracy can be improved by ATA-Module using the information that enhances the evidence of being a fake as determined by the discriminator.

Adversarial loss is one of the influences from the discriminator to the generator, but it is not enough to transfer the information in discriminator to the generator. This is because adversarial loss only transmits the result of whether the discriminator was deceived or not to the generator as a loss function. It is not possible to know the regions that the discriminator was not deceived. By using the ATA-Module, the pixels that were not deceived can be transmitted to the generator in a precise and efficient manner.

As show in Figure~\ref{fig:ATA-Module}, we feed the feature maps in discriminator into $1\times1$ convolution layers to generate new feature maps \textit{\textbf{Query}} and \textit{\textbf{Key}}, respectively. We are inspired by Self-Attention GAN (SAGAN)~\cite{sagan} to reduce the channel number to $C/8$ for memory efficiency. Then, we reshape them to $C/8 \times (H \times W)$. After we perform a matrix multiplication between the transpose of \textit{\textbf{Query}} and \textit{\textbf{Key}}, and we use a softmax function to calculate an attention map as 
\begin{equation}
\label{eq:Attention_weight}
w_{ij}=\frac{\exp({\textit{\textbf{Query}}}_{i}^T ~{\textit{\textbf{Key}}}_{j})}{\sum_{j=1}^{H \times W} {\exp({\textit{\textbf{Query}}}_{i}^T~{\textit{\textbf{Key}}}_{j})}},
\end{equation}
where $w_{ij}$ measures the $j^{th}$ \textit{\textbf{Query}}'s impact on $i^{th}$ \textit{\textbf{Key}}. If two pixels belong to the same class, the feature representation will be similar. $H \times W$ is the total number of pixels.

Meanwhile, we feed the feature map in discriminator into $1\times1$ convolution layer to generate a new feature map \textit{\textbf{Value}} and reshape it to $C/2 \times (H \times W)$. Then, we perform a matrix multiplication between the attention map and the transpose of \textit{\textbf{Value}} and reshape the result to $C/2 \times H \times W$, and then to $C \times H \times W$ by $1 \times 1$ convolution. Finally, we multiply it by a scale parameter $\alpha$ and perform a element-wise sum operation with the feature maps in generator to obtain the final output as
\begin{equation}
\label{eq:Attention_matual}
S_i=\alpha \sum_{j=1}^{H \times W}{(w_{ij}~ \textit{\textbf{Value}}_j^T)^T+F_i},
\end{equation}
where $\alpha$ is initialized as 0 and gradually trains to assign more weight~\cite{sagan}. $S_i$ indicates the output and $F_i$ indicates the feature map of the generator.

Equation~\ref{eq:Attention_matual} means that the output $S_i$ is the weighted sum of all positions in the discriminator. Therefore, the difference between the real and fake extracted by discriminator is aggregated by the similarity of each pixel.
The segmentation accuracy is improved by transmitting the information of discriminator to the generator.

\begin{figure}[t] 
    \centering
    \includegraphics[width=0.95\linewidth]{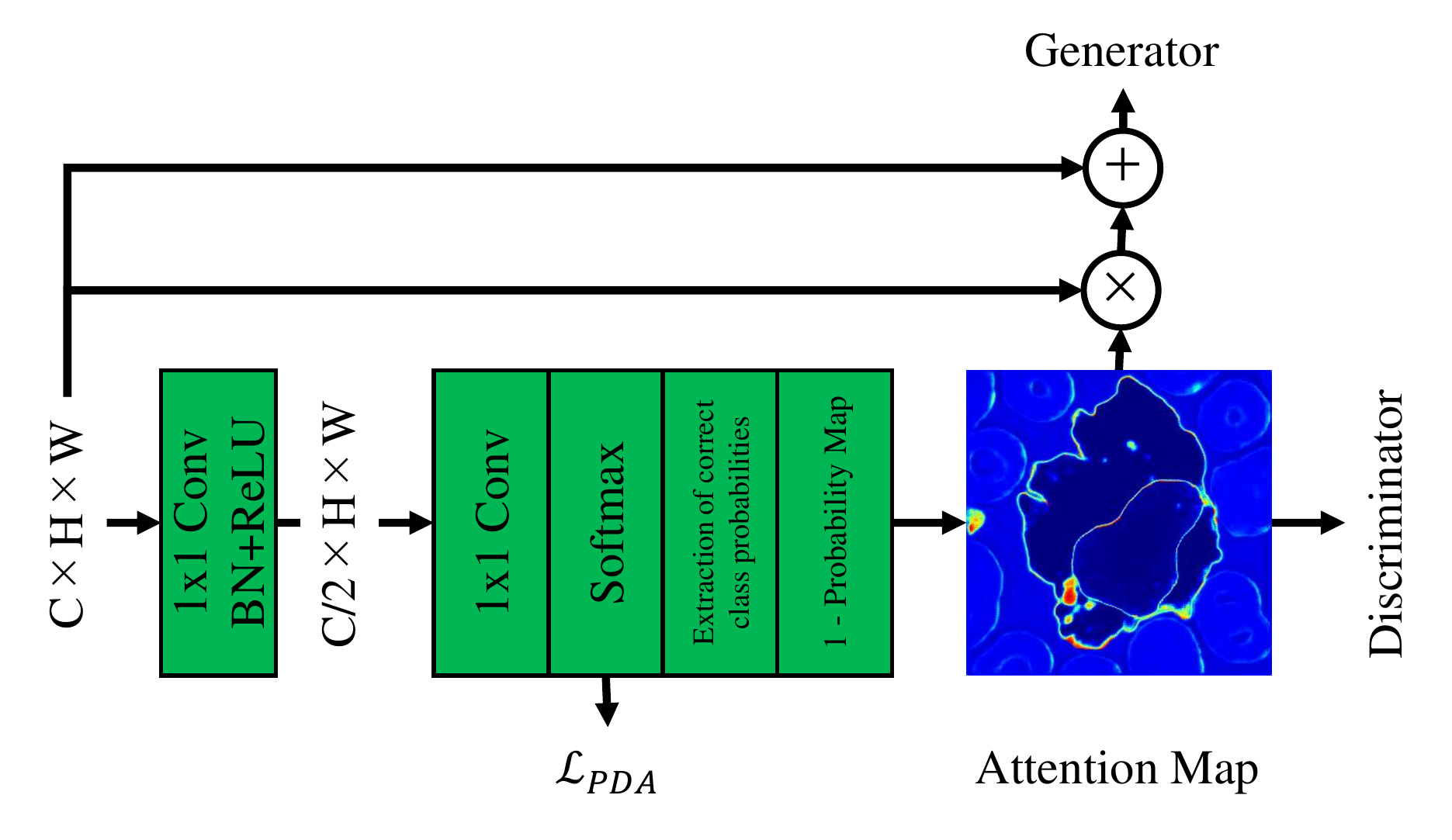}
  \caption{\textbf{Network structure of Top-Down PDA-Module.} The predicted probability of true class at the decoder is subtracted from 1 to form an attention map. Pixels with low probability have high attention values. Attention map is used for both of generator and  discriminator.}
  \label{fig:Top-Down PDA-Module}
\end{figure}

\subsection{Top-Down PDA Module}

The discriminator trains the difference between the generated image and ground truth to prevent the generator from deceiving the generated image as the real one. By leaking the information about the difficult pixels from the generator to the discriminator, the discriminator can use the information of the difficult pixels in the generator as a reason to judge the difference between the generated image and ground truth, thereby improving the performance of the discriminator. As the performance of the discriminator is improved, the performance of the generator is also improved by training the generator to fool the discriminator by adversarial loss. 
Therefore, we propose the Top-Down Pixel-wise Difficulty Attention Module (Top-Down PDA-Module), which creates an attention map that enhances the pixels with high difficulty in the generator and leaks the attention map to the discriminator. Top-Down PDA-Module is a top-down attention mechanism that refers to ground truth, so it works only during training.
Since we cannot use to ground truth during inference, we do not use Top-Down PDA-Module in test phase.
This is not a problem because the pixel-wise difficulty is close to 0 throughout training.

In semantic segmentation, the output layer compresses the number of channels of the feature map into the number of classes, and converts them into probability for each class by a softmax function. In this case, the pixels with low probability for true class can be considered difficult. This information is explicitly given to the discriminator as an attention map, and it can be used as a reason to judge the difference between the generated image and the ground truth, so that the discriminator can accurately classify.

As show in Figure~\ref{fig:Top-Down PDA-Module}, Top-Down PDA-Module performs segmentation at each resolution of the decoder in the generator in order to enhance the pixels with high difficulty in classification based on the probability. This makes it possible to calculate the difficulty level for each pixel at each resolution. For a feature map $A \in \mathbb{R}^{C \times H \times W}$ in the decoder, the number of channels is compressed to $B \in \mathbb{R}^{C/2 \times H/W}$ by $1 \times 1$ convolution, Batch Normalization and ReLU function. Then we compress the channel dimension to the number of classes by $1 \times 1$ convolution, and we perform upsampling to the original input image size and use the softmax function to obtain the probability values for all pixels. From this probability map, we pick up the probability of true class at each pixel to create a map $C \in \mathbb{R}^{H_{in} \times W_{in}}$. 
By subtracting the map from 1 (true probability) and downsampling to the original feature map $C '\in\mathbb{R}^{H \times W}$, we can create an attention map  with high values for difficult pixels.

Attention Map created from the correct answer probability map is multiplied by the feature map from the decoder to enhance the pixels with high difficulty. Furthermore, by performing a residual addition with the feature map from the decoder, the gradient does not disappear and the magnitude of the value does not change drastically between training and inference. In addition, by calculating the loss function using the probability map in the middle of Top-Down PDA-Module, the training process can be optimized by back propagation of error like auxiliary loss from the middle output. Since the size of the probability map is the same as that of the input image, the training process is close to the final output even with low-resolution feature maps. By using the segmentation output in the middle layer of the decoder, the Top-Down PDA-Module is able to create an attention map that focuses on the difficulty of pixels. Softmax cross-entropy is used as a loss function.

The decoder's feature map $A$ in the generator can be multiplied by the attention map, and $A$ is enhanced based on the difficulty level by the residual addition. In addition, we also leak the created attention map to the discriminator. Thus, the attention map is used for both of generator and discriminator.
In conventional adversarial training, the difference between the generated image and ground truth, which is the reason for the discriminator's decision, is acquired bottom-up manner during the training process, so that unintended pixels with low importance may become the basis for the decision. 
If the discriminator clearly trains the difference between the generated image and ground truth based on the attention map and the judged results are used for training the generator, 
the accuracy can be improved from the case of the generator alone. Attention Map is multiplied with the feature map at the same resolution in the discriminator, and the pixels with high difficulty by the generator are enhanced. Furthermore, by performing a residual addition, the discriminator can leak the information that the generator is training to be important.

\begin{figure*}[t] 
    \centering
    \includegraphics[width=0.89\linewidth]{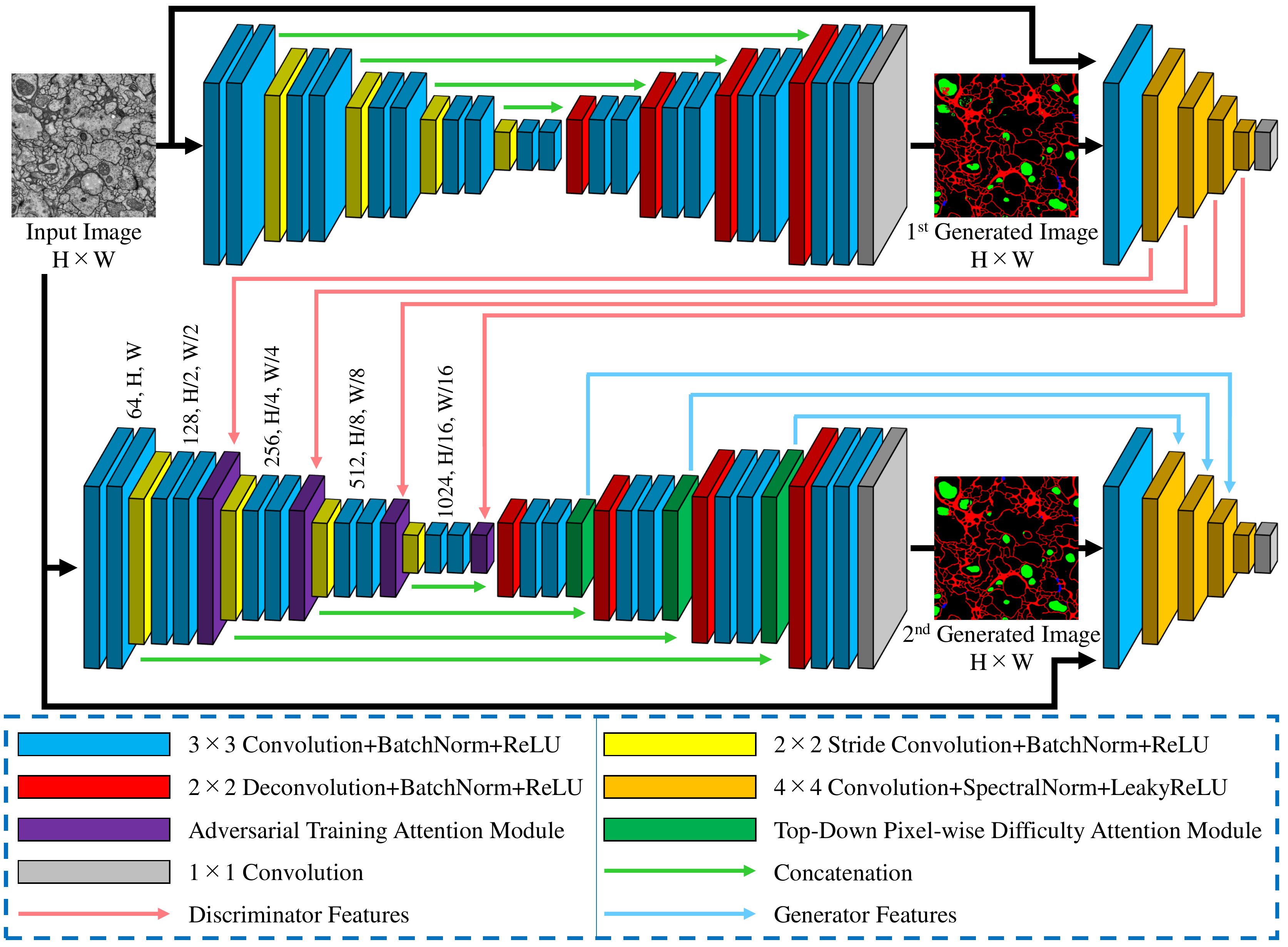}
  \caption{\textbf{Network structure of Adversarial Mutual Network.} First, the generator segments an image without attention mechanism. The segmentation result is fed into the discriminator. An attention map is created by ATA-Module from the discriminator's feature map, and it is used in the second generator. Next, an attention map is created from the decoder in the generator by Top-Down PDA-Module, and it is leaked to the generator and the discriminator. Finally, the discriminator judges true or false. The first and second generators and discriminators share the same weights except for the attention mechanism.}
  \label{fig:AML-Net}
\end{figure*}

\subsection{Adversarial Mutual Leakage Network}

The performance of the generator is improved by using our proposed two modules; the ATA-Module which leaks from the discriminator to the generator and the Top-Down PDA-Module which leaks from the generator to the discriminator. The performance of the generator can be further improved by using both of our proposed modules. Finally, we propose Adversarial Mutual Leakage Network (AML-Net), which combines ATA-Module and Top-Down PDA-Module to leak the information in both directions.

Whole structure is shown in Figure~\ref{fig:AML-Net}. The generator is based on U-Net and the discriminator is a CNN with 6 convolutional layers. AML-Net performs segmentation twice using the generators with shared weights for ATA-Module, which uses the feature map of the discriminator. First, the segmentation is done by the generator without ATA-Module and Top-Down PDA-Module, and then the result is fed into the discriminator to classify real or fake. The feature map in the discriminator is obtained to use for ATA-Module. The attention map made from feature map in the discriminator at each resolution in discriminating the generated image at the first round is sent to the encoder at the same resolution of the generator at the second round through ATA-Module. This allows us to generate an image that is closer to ground truth than the generated image at the first round. 

Next, the decoder part of each resolution at the second round generator uses the Top-Down PDA-Module to create an attention map based on the difficulty level in the generator. This allows the generator to accurately train the pixels of high difficulty that are difficult to predict using explicit information from ground truth. The created attention map is also leaked to the discriminator to improve the discrimination performance between the generated image and ground truth, and the generator trains to successfully deceive the discriminator by adversarial loss. 

Discriminator in AML-Net has an encoder structure and extracts features from the input image paired with the generated image or ground truth by $3\times3$ convolution. 
Next, we perform downsampling by convolution with a stride number of 2 using a kernel of $4 \times 4$ at each resolution. 
In this process, we use Spectral Normalization~\cite{miyato2018spectral}, which is a GAN training stabilization method. We apply Leaky ReLU to the obtained feature map, and extract features by multiplication and residual addition with the attention map leaked from the generator. In the final layer, a $1 \times 1$ convolution is used to perform binary classification so that the output is true for ground truth and fake for the generated image. 
In addition, by referring to the Patch GAN proposed by pix2pix~\cite{pix2pix}, the discriminator uses $16 \times 16$ pixels of the input image as a patch to determine whether the image is real or fake. This improves the generator based on local judgments in the cropped patch instead of global judgments for a single image. AML-Net trains the generator with the loss function used for segmentation and the adversarial loss function. 
In final output of the U-Net used in the generator, the loss (${\cal L}_{CE}$) is calculated using softmax cross-entropy. For Top-Down PDA-Module at each of the three resolutions included in the generator, softmax cross-entropy is also used to calculate the respective losses (${\cal L}_{PDA1}$, ${\cal L}_{PDA2}$, and ${\cal L}_{PDA3}$). In addition, by using adversarial loss (${\cal L}_{adv}$) to determine whether the generator deceives the discriminator, the generator is trained by considering the realness that cannot be determined only by the loss of the generator (${\cal L}_{CE}$). Adversarial loss ${\cal L}_{adv}$ is given as
\begin{eqnarray}
\label{eq:adversarial_loss}
{\cal L}_{adv}(G,D) = \mathbb{E}_{x,y\sim p_{data}(x,y)}\left [ logD(x,y) \right ] \nonumber \\ 
+\mathbb{E}_{x\sim p_{data}(x)}\left [ log(1-D(x,G(x))) \right ].
\end{eqnarray}
There are two inputs in Equation~\ref{eq:adversarial_loss}, $x$ represents the input image and $y$ represents ground truth. In addition, $p_{data}$ represents the distribution of training data, and $x,y~p_{data (x,y)}$ is the process of sampling data from $p_{data}$. Let $G(x)$ denote the segmentation result output by the generator, and $D(\centerdot)$ denote the result output by the discriminator, which determines whether the data is real or fake. Based on this equation, adversarial training is optimized with the following Min-Max objective function. $G(x)$ denotes the segmentation result output by the generator, and $D(\centerdot)$ denotes the discriminant result output by the discriminator. From Equation~\ref{eq:adversarial_loss}, adversarial training is optimized with the Min-Max objective function as
\begin{eqnarray}
\label{eq:min_max}
\,arg\, \underset{G}{\min}\,\underset{D}{\max}{\cal L}_{adv}(G,D).
\end{eqnarray}
The generator $G$ minimizes ${\cal L}_{adv}$ so that the segmentation image $G(x)$ deceives the discriminator, and the discriminator $D$ maximizes ${\cal L}_{adv}$ so that the output $G(x)$ of the generator is fake and the ground truth $y$ is true. Especially, the adversarial loss for the generator is as follows:
\begin{eqnarray}
\label{eq:adversarial_G}
{\cal L}_{adv}(G)=\mathbb{E}_{x\sim p_{data}(x)}\left [ log(1-D(x,G(x))) \right ]
\end{eqnarray}

The final loss function of the generator is as
\begin{eqnarray}
{\cal L}_{total}={\cal L}_{CE}+{\cal L}_{PDA1}+{\cal L}_{PDA2} \nonumber \\
+{\cal L}_{PDA3}+\lambda_{adv}{\cal L}_{adv}(G),
\label{eq:total_loss}
\end{eqnarray}
where $\lambda_{adv}$ is a hyperparameter, which is set weakly to prevent the discriminator from becoming too strong. In this paper, we set $\lambda_{adv}=0.01$ as same as pixe2pix.


\section{Experiments}

In this section, we first introduce the implementation details and the dataset used in our experiments. Next, we evaluated our method on two cell image datasets. 

\subsection{Implementation Details}

We adopted Adam as the Optimizer in our experiments. The training coefficients were set to $\alpha=10^{-3}$, $\beta_{1}=0.9$, $\beta_{2}=0.9$, and $\varepsilon=10^{-8}$. A single GeForce RTX 2080 Ti GPU was used for training in this experiment. 5-fold cross-validation were conducted to properly evaluate the generalization performance.
The model given the highest mean Intersection of Union (mIoU) for the validation images was used for the evaluation. Since the experimental results are changed according to the initial seed of a random function, the same experiment was conducted three times with different initial values, and the results were evaluated based on the average of 15 times in total.

\subsection{Dataset}

\subsubsection{White Blood Cell}

White Blood Cell(WBC)~\cite{zheng2018wbc} consists of 100 images of $300 \times 300$ pixels in three classes: cell nucleus of white blood cells, cytoplasm, and background including red blood cells. For the experiment, the images were resized to $320 \times 320$ pixels by bilinear interpolation. 64 images were training images, 16 images were validation images, and 20 images were test images. The batch size is set to 2 for training. In the segmentation images, white color indicates cell nucleus, gray color indicates cytoplasm, and black color indicates the background including red blood cells.

\subsubsection{Drosophila Cell Image}

Drosophila cell image~\cite{sstem} consists of 20 images including 4 classes of cell membrane, cytoplasm, mitochondria, and synapses with an image size of $1024 \times 1024$ pixels. In experiment, we divided the images into 12 training images, 3 validation images, and 5 test images, and divided the original image size into 16 regions of $256 \times 256$ pixels without overlap. The final number of images was 192 for training, 48 for validation, and 60 for evaluation. The batch size is set to 4 for training. In the segmentation images, red indicates cell membrane, black indicates cell membrane, green indicates mitochondria, and blue indicates synapses.

\begin{table}[t]
    \centering
    \caption{IoU Accuracy on WBC dataset}
    \fontsize{5.5pt}{8pt}\selectfont
    \label{table:WBC_IoU}
    \begin{tabular}{l|cccc}
    \hline
Method
& Cytoplasm~[\%]                 & Nucleus~[\%]                   & Background~[\%]                & Mean IoU~[\%] \\ \hline \hline
U-Net\cite{unet}
& 69.95±7.89                          & 89.31±2.06                          & 94.42±2.49                          & 84.56±3.86                                 \\

Attention U-Net\cite{oktay2018attention_unet}
& 70.48±8.51                          & 88.42±2.65                          & 94.95±2.46                          & 84.62±4.15                                 \\

pix2pix\cite{pix2pix}
& 71.62±5.56                          & 89.45±1.88                          & 95.21±1.74                          & 85.43±2.70                                 \\

SAGAN\cite{sagan}
& 74.32±4.92                          & {\color{green} \textbf{89.94}}±1.37 & 95.87±1.50                        & 86.71±2.35        \\

PDA-Module\cite{matsuzuki2019PDA}
& 72.80±10.13                         & 88.40±4.19                          & 95.64±3.23                          & 85.62±5.45                                 \\

Deeplabv3+\cite{deeplabv3plus}
& 78.52±4.35                          & {\color{red} \textbf{91.51}}±0.97   & {\color{green} \textbf{97.07}}±0.91 & {\color{green} \textbf{89.03}}±1.89  \\

FastFCN\cite{wu2019fastfcn}
& {\color{red} \textbf{81.52}}±2.62   & 89.25±1.76                         & {\color{red} \textbf{98.22}}±0.34    & {\color{blue} \textbf{89.66}}±1.47  \\ \hline

ATA-Module
& 69.12±7.65                          & 89.31±1.86                          & 94.04±2.70                          & 84.16±3.82                                 \\

Top-Down PDA-Module
& {\color{green} \textbf{77.40}}±7.70 & 89.84±2.40                          & 96.76±2.09                            & 88.00±3.73 \\

AML-Net
& {\color{blue} \textbf{81.12}}±5.02  & {\color{blue} \textbf{90.81}}±1.56   & {\color{blue} \textbf{97.59}}±1.26   & {\color{red} \textbf{89.84}}±2.41 \\ \hline

    \end{tabular}\\~\\

    \caption{Precision-Recall Accuracy on WBC dataset}
    \fontsize{5.5pt}{8pt}\selectfont
    \label{table:WBC_PR}
    \begin{tabular}{l|ccc}
    \hline
\multicolumn{1}{c|}{Precision}
& \multirow{2}{*}{Cytoplasm~[\%]}                 & \multirow{2}{*}{Nucleus~[\%]}                   & \multirow{2}{*}{Background~[\%]} \\
\multicolumn{1}{c|}{Recall}&&& \\ \hline \hline
\multirow{2}{*}{U-Net\cite{unet}}
& 76.30±8.51                          & 92.47±2.81 & {\color{blue} \textbf{99.28}}±0.30 \\
& {\color{green} \textbf{89.26}}±2.56 & 96.42±2.29                          & 95.07±2.62  \\ \hline

\multirow{2}{*}{Attention U-Net\cite{oktay2018attention_unet}}
& 77.95±8.15                          & 91.86±3.35                          & 99.11±0.45  \\
& 87.71±4.81                          & 96.03±2.26                          & 95.77±2.36  \\ \hline

\multirow{2}{*}{pix2pix\cite{pix2pix}}
& 79.66±6.72                          & 92.38±1.99                          & 99.03±0.48  \\
& 87.82±3.16                          & 96.64±2.54                          & 96.13±2.01  \\ \hline

\multirow{2}{*}{SAGAN\cite{sagan}}
& 81.74±6.21                          & {\color{green} \textbf{92.81}}±1.55  & 99.21±0.20  \\
& {\color{blue} \textbf{89.27}}±2.26  & 96.71±1.87                          & 96.61±1.60  \\ \hline

\multirow{2}{*}{PDA-Module\cite{matsuzuki2019PDA}}
& 81.79±10.79                         & 90.80±5.23                          & 99.23±0.38  \\
& 86.88±6.15                          & {\color{green} \textbf{97.25}}±1.72                          & 96.37±3.43  \\ \hline

\multirow{2}{*}{Deeplabv3+\cite{deeplabv3plus}}
& {\color{blue} \textbf{90.18}}±3.09 & {\color{red} \textbf{94.17}}±1.24                          & 98.32±0.72  \\
& 85.82±3.64                          & 97.02±0.53  & {\color{green} \textbf{98.70}}±0.63  \\ \hline

\multirow{2}{*}{FastFCN+\cite{wu2019fastfcn}}
& {\color{red} \textbf{92.90}}±2.05 & 91.68±2.04                          & 99.02±0.38  \\
& 86.99±3.05                          & 97.14±1.13  & {\color{red} \textbf{99.19}}±0.40  \\ \hline
 \hline

\multirow{2}{*}{ATA-Module(ours)}
& 75.96±10.37                         & 92.37±2.70                          & {\color{green} \textbf{99.27}}±0.34  \\
& 89.15±3.79                          & 96.54±2.56                          & 94.71±2.95  \\ \hline

\multirow{2}{*}{Top-Down PDA-Module(ours)}
& 85.81±7.88  & 92.16±2.62                          & 99.24±0.29 \\
& 88.66±3.88                          & {\color{blue} \textbf{97.32}}±1.77 & 97.48±2.08 \\ \hline

\multirow{2}{*}{AML-Net(ours)}
& {\color{green} \textbf{89.28}}±6.04   & {\color{blue} \textbf{92.93}}±1.51   & {\color{red} \textbf{99.33}}±0.14 \\
& {\color{red} \textbf{89.94}}±1.42   & {\color{red} \textbf{97.57}}±1.38   & {\color{green} \textbf{98.24}}±1.32 \\ \hline
    \end{tabular}
    \normalsize 
\end{table}

\begin{figure*}
\centering
  \begin{minipage}[b]{.15\linewidth}
    \includegraphics[width=\columnwidth]{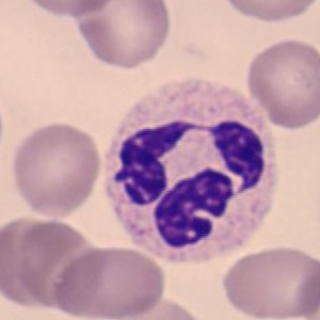}
    \centerline{Input}
  \end{minipage}%
  ~
  \begin{minipage}[b]{.15\linewidth}
    \centering
    \includegraphics[width=\columnwidth]{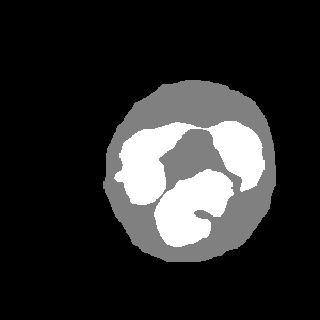}
    \centerline{Ground Truth}
  \end{minipage}%
  ~
  \begin{minipage}[b]{.15\linewidth}
    \centering
    \includegraphics[width=\columnwidth]{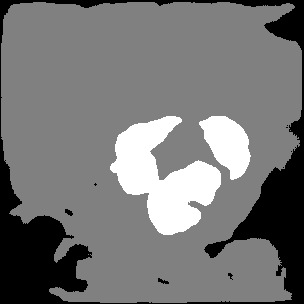}
    \centerline{U-Net\cite{unet}}
  \end{minipage}%
  ~
  \begin{minipage}[b]{.15\linewidth}
    \centering
    \includegraphics[width=\columnwidth]{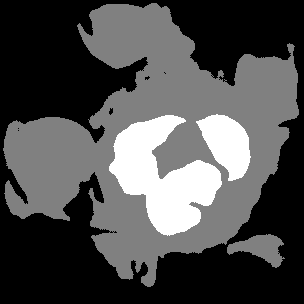}
    \centerline{Attention U-Net\cite{oktay2018attention_unet}}
  \end{minipage}
  ~
  \begin{minipage}[b]{.15\linewidth}
    \centering
    \includegraphics[width=\columnwidth]{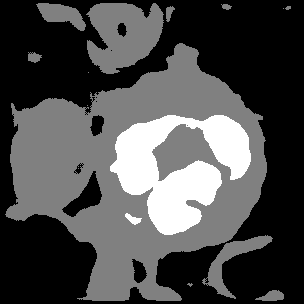}
    \centerline{pix2pix\cite{pix2pix}}
  \end{minipage}%
  ~
  \begin{minipage}[b]{.15\linewidth}
    \centering
    \includegraphics[width=\columnwidth]{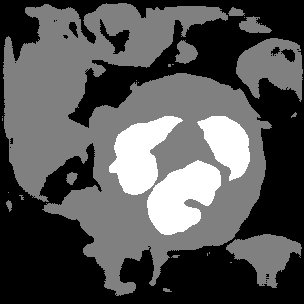}
    \centerline{SAGAN\cite{sagan}}
  \end{minipage}%
  \\~\\
  \begin{minipage}[b]{.15\linewidth}
    \centering
    \includegraphics[width=\columnwidth]{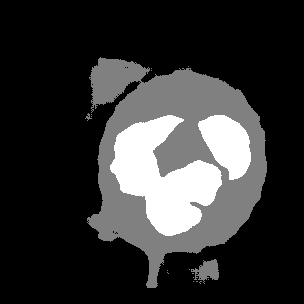}
    \centerline{PDA-Module\cite{matsuzuki2019PDA}}
    \centerline{}
    \centerline{}
  \end{minipage}%
  ~
  \begin{minipage}[b]{.15\linewidth}
    \centering
    \includegraphics[width=\columnwidth]{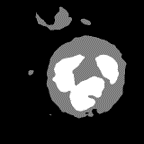}
    \centerline{Deeplabv3+\cite{deeplabv3plus}}
    \centerline{}
    \centerline{}
  \end{minipage}%
  ~
  \begin{minipage}[b]{.15\linewidth}
    \centering
    \includegraphics[width=\columnwidth]{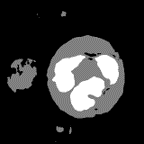}
    \centerline{FastFCN\cite{wu2019fastfcn}}
    \centerline{}
    \centerline{}
  \end{minipage}%
  ~
  \begin{minipage}[b]{.15\linewidth}
    \centering
    \includegraphics[width=\columnwidth]{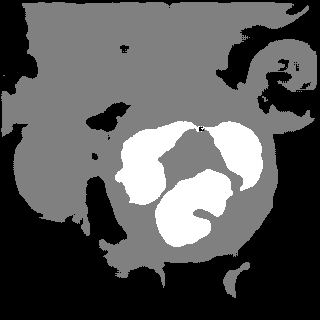}
    \centerline{ATA-Module}
    \centerline{}
    \centerline{}
  \end{minipage}%
  ~
  \begin{minipage}[b]{.15\linewidth}
    \centering
    \includegraphics[width=\columnwidth]{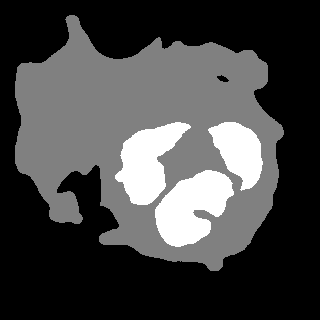}
    \centerline{Top-Down}
    \centerline{PDA-Module}
    \centerline{}
  \end{minipage}%
  ~
  \begin{minipage}[b]{.15\linewidth}
    \centering
    \includegraphics[width=\columnwidth]{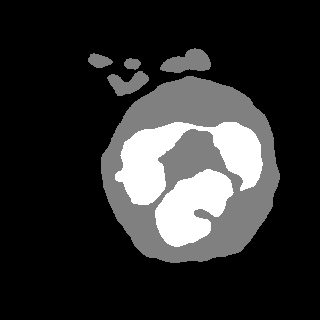}
    \centerline{AML-Net}
    \centerline{}
    \centerline{}
  \\
  \end{minipage}

\centering
  \begin{minipage}[b]{.15\linewidth}
    \includegraphics[width=\columnwidth]{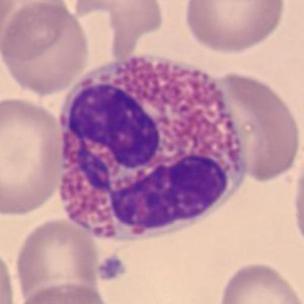}
    \centerline{Input}
  \end{minipage}%
  ~
  \begin{minipage}[b]{.15\linewidth}
    \centering
    \includegraphics[width=\columnwidth]{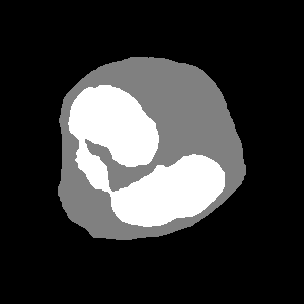}
    \centerline{Ground Truth}
  \end{minipage}%
  ~
  \begin{minipage}[b]{.15\linewidth}
    \centering
    \includegraphics[width=\columnwidth]{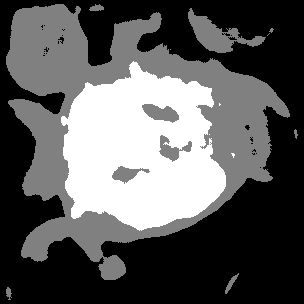}
    \centerline{U-Net\cite{unet}}
  \end{minipage}%
  ~
  \begin{minipage}[b]{.15\linewidth}
    \centering
    \includegraphics[width=\columnwidth]{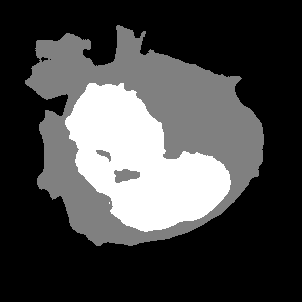}
    \centerline{Attention U-Net\cite{oktay2018attention_unet}}
  \end{minipage}
  ~
  \begin{minipage}[b]{.15\linewidth}
    \centering
    \includegraphics[width=\columnwidth]{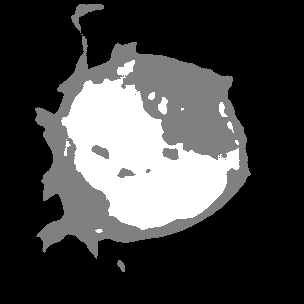}
    \centerline{pix2pix\cite{pix2pix}}
  \end{minipage}%
  ~
  \begin{minipage}[b]{.15\linewidth}
    \centering
    \includegraphics[width=\columnwidth]{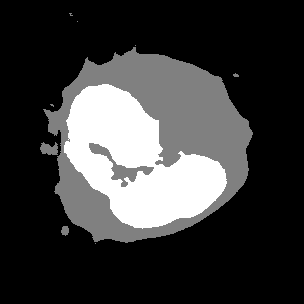}
    \centerline{SAGAN\cite{sagan}}
  \end{minipage}%
  \\~\\
  \begin{minipage}[b]{.15\linewidth}
    \centering
    \includegraphics[width=\columnwidth]{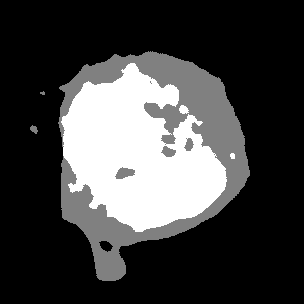}
    \centerline{PDA-Module\cite{matsuzuki2019PDA}}
    \centerline{}
  \end{minipage}%
  ~
  \begin{minipage}[b]{.15\linewidth}
    \centering
    \includegraphics[width=\columnwidth]{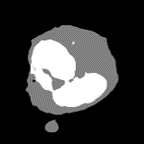}
    \centerline{Deeplabv3+\cite{deeplabv3plus}}
    \centerline{}
  \end{minipage}%
  ~
  \begin{minipage}[b]{.15\linewidth}
    \centering
    \includegraphics[width=\columnwidth]{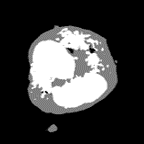}
    \centerline{FastFCN\cite{wu2019fastfcn}}
    \centerline{}
  \end{minipage}%
  ~
  \begin{minipage}[b]{.15\linewidth}
    \centering
    \includegraphics[width=\columnwidth]{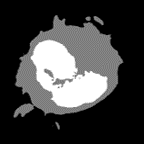}
    \centerline{ATA-Module}
    \centerline{}
  \end{minipage}%
  ~
  \begin{minipage}[b]{.15\linewidth}
    \centering
    \includegraphics[width=\columnwidth]{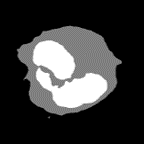}
    \centerline{Top-Down}
    \centerline{PDA-Module}
  \end{minipage}%
  ~
  \begin{minipage}[b]{.15\linewidth}
    \centering
    \includegraphics[width=\columnwidth]{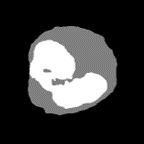}
    \centerline{AML-Net}
    \centerline{}
  \\
  \end{minipage}
  \caption{Segmentation result on WBC dataset}
  \label{fig:WBC_Outputs}
\end{figure*}

\subsection{Experiments Results}

\subsubsection{Results on WBC dataset}

We use U-Net~\cite{unet} as a baseline, which has been widely applied in medical and biological images. Following methods are the comparison methods. two adversarial training methods; pix2pix~\cite{pix2pix} based on U-Net with a discriminator and Self-Attention GAN (SAGAN)~\cite{sagan} which adds Self-Attention~\cite{transformer} to the discriminator of pix2pix. We also evaluate attention U-Net~\cite{oktay2018attention_unet} introduced an attention mechanism called sSE-Block~\cite{roy2018scsenet} to Skip Connection of U-Net. PDA-Module~\cite{matsuzuki2019PDA} is a method that applies a bottom-up attention mechanism based on difficulty to U-Net. 

Table~\ref{table:WBC_IoU} shows the IoU accuracy on the WBC dataset, and Table~\ref{table:WBC_PR} shows the experimental results of evaluating WBC dataset with Precision and Recall. Bold red letters in Table represent the best accuracy, and blue and green bold letters are the second and third best. 
We see that the accuracy of AML-Net is better than that of conventional methods based on U-Net. In particular, we confirm that the accuracies of Top-Down PDA-Module and AML-Net were improved in almost classes compared to pix2pix and SAGAN, which are adversarial training methods without leakage. Especially, in cytoplasm which is the most difficult, large accuracy improvement was observed. In addition, the accuracy of AML-Net was better than other conventional methods in terms of the mean and standard deviation. This indicates that the accuracy of AML-Net is stable and does not depend on the combination of datasets or initial values. In addition, we confirm that the mean IoU of AML-Net based on U-Net is higher than that of Deeplabv3+~\cite{deeplabv3plus} and FastFCN~\cite{wu2019fastfcn} based on ResNet-50~\cite{resnet}. By avoiding the use of very deep networks such as ResNet-50, the training and inference speed can be accelerated, which is an advantage for cell biologists in practical use. From Table~\ref{table:WBC_PR}, we can see that AML-Net has the best Recall for cytoplasm and cell nucleus, and Precision also shows high accuracy. Conventional methods only have high precision or high recall, but AML-Net has an advantage in the balance of precision and recall.

Figure~\ref{fig:WBC_Outputs} shows segmentation results. We see that the most of conventional methods in the upper image group over-detect the cytoplasm because the color of red blood cells and cytoplasm are similar. In addition, in the lower image group, we can see that the cytoplasm staining is darker and therefore closer to the color of the cell nucleus, and that segmentation is not working well with many conventional methods. Our AML-Net can accurately segment the regions even for images with very similar colors.

\begin{table}[t]
    \centering
    \caption{Accuracy on Drosophila dataset}
    \fontsize{5.5pt}{8pt}\selectfont
    \label{table:drosophila_IoU}
    \begin{tabular}{l|ccccc}
    \hline
Method                    & \rotatebox[origin=c]{90}{Membrane~[\%]}  & \rotatebox[origin=c]{90}{ Mitochondria~[\%] }  & \rotatebox[origin=c]{90}{Synapse~[\%]} & \rotatebox[origin=c]{90}{Cytoplasm~[\%]} & \rotatebox[origin=c]{90}{Mean IoU~[\%]} \\ \hline \hline
U-Net\cite{unet}
& 73.76±2.23                         & 69.66±7.21                         & 42.84±3.71                         & 91.96±0.50                         & 69.56±3.22  \\

Attention U-Net\cite{oktay2018attention_unet}
& 76.26±1.18                         & 76.53±2.97                         & 44.46±3.33                         & 92.51±0.23                         & 72.44±1.43  \\

pix2pix\cite{pix2pix}
& 75.80±0.85                         & 76.55±2.30                         & 43.59±3.27                         & 92.39±0.26                         & 72.08±1.09  \\

SAGAN\cite{sagan}
& {\color{blue} \textbf{76.38}}±0.63 & 78.49±1.70                         & 43.16±2.95                         & {\color{blue} \textbf{92.64}}±0.12 & 72.66±1.10  \\

PDA-Module\cite{matsuzuki2019PDA}
& 75.02±1.64                         & 75.17±2.20                         & 45.01±3.28                         & 92.19±0.30                         & 71.85±1.15  \\

Deeplabv3+\cite{deeplabv3plus}
& 64.75±2.10                           & 45.67±7.77                        & 36.18±4.12                         & 90.08±0.36                        & 59.17±3.11  \\

FastFCN\cite{wu2019fastfcn}
& 74.93±0.58                         & {\color{blue} \textbf{79.26}}±1.82  & {\color{green} \textbf{48.11}}±2.91                         & 92.19±0.25                         & {\color{green} \textbf{73.62}}±1.25  \\ \hline

ATA-Module
& {\color{red} \textbf{77.64}}±0.49  & {\color{green} \textbf{78.92}}±1.77 & 47.15±4.01                     & {\color{red} \textbf{92.82}}±0.21  & {\color{blue} \textbf{74.13}}±1.31 \\

Top-Down PDA-Module
& 75.42±0.95                         & 74.78±2.59                         & {\color{red} \textbf{49.48}}±3.43  & 92.03±0.29                         & 72.93±1.40  \\

AML-Net
& {\color{green} \textbf{76.27}}±0.42  & {\color{red} \textbf{79.84}}±1.15  & {\color{blue} \textbf{49.08}}±3.17 & {\color{green} \textbf{92.57}}±0.21& {\color{red} \textbf{74.44}}±0.80 \\ \hline
    \end{tabular}
    \normalsize 
\end{table}

\begin{figure*}
\centering
  \begin{minipage}[b]{.15\linewidth}
    \includegraphics[width=\columnwidth]{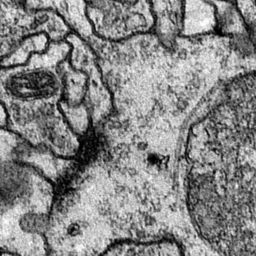}
  \end{minipage}%
  ~
  \begin{minipage}[b]{.15\linewidth}
    \centering
    \includegraphics[width=\columnwidth]{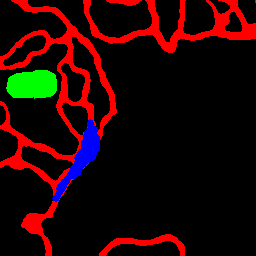}
  \end{minipage}%
  ~
  \begin{minipage}[b]{.15\linewidth}
    \centering
    \includegraphics[width=\columnwidth]{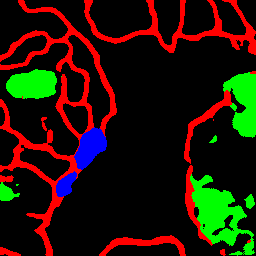}
  \end{minipage}%
  ~
  \begin{minipage}[b]{.15\linewidth}
    \centering
    \includegraphics[width=\columnwidth]{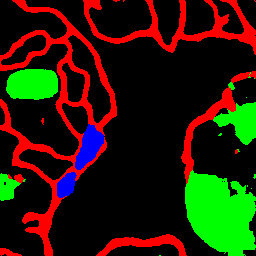}
  \end{minipage}%
  ~
  \begin{minipage}[b]{.15\linewidth}
    \centering
    \includegraphics[width=\columnwidth]{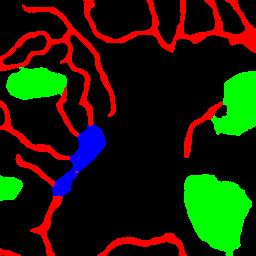}
  \end{minipage}%
  ~
  \begin{minipage}[b]{.15\linewidth}
    \centering
    \includegraphics[width=\columnwidth]{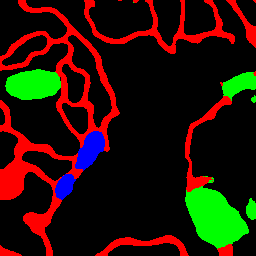}
  \end{minipage}
  \\~\\
  \begin{minipage}[b]{.15\linewidth}
    \includegraphics[width=\columnwidth]{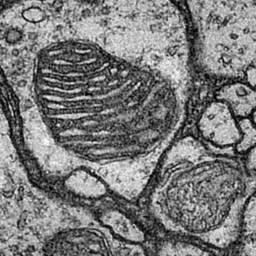}
    \centerline{Input}
    \centerline{}
  \end{minipage}%
  ~
  \begin{minipage}[b]{.15\linewidth}
    \centering
    \includegraphics[width=\columnwidth]{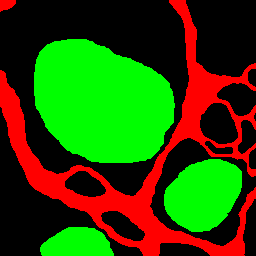}
    \centerline{Ground Truth}
    \centerline{}
  \end{minipage}%
  ~
  \begin{minipage}[b]{.15\linewidth}
    \centering
    \includegraphics[width=\columnwidth]{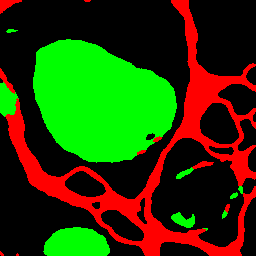}
    \centerline{U-Net\cite{unet}}
    \centerline{}
  \end{minipage}%
  ~
  \begin{minipage}[b]{.15\linewidth}
    \centering
    \includegraphics[width=\columnwidth]{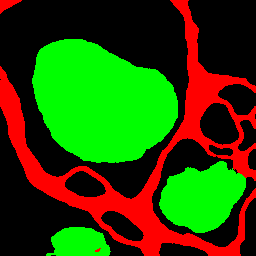}
    \centerline{ATA-Module}
    \centerline{}
  \end{minipage}%
  ~
  \begin{minipage}[b]{.15\linewidth}
    \centering
    \includegraphics[width=\columnwidth]{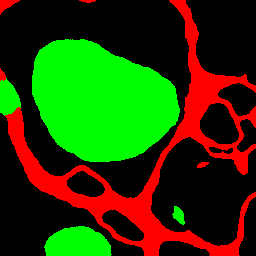}
    \centerline{Top-Down}
    \centerline{PDA-Module}
  \end{minipage}%
  ~
  \begin{minipage}[b]{.15\linewidth}
    \centering
    \includegraphics[width=\columnwidth]{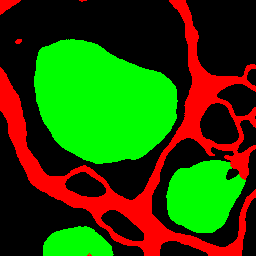}
    \centerline{AML-Net}
    \centerline{}
  \end{minipage}
  \caption{Segmentation results on Drosophila dataset}
  \label{fig:drosophila_Outputs}
\end{figure*}

\subsubsection{Results on Drosophila Cell Images}

As shown in Table~\ref{table:drosophila_IoU}, the proposed method improved the accuracy compared to the conventional methods in many classes. In particular, our proposed AML-Net and Top-Down PDA-Module improved the accuracy of synapses, which is the most difficult class. Top-Down PDA-Module is a top-down attention mechanism using ground-truth, and it is considered that the accuracy is improved due to its ability to create attention maps that explicitly strengthen particularly difficult synapses.
In addition, the accuracy of ATA-Module was improved in many classes compared to conventional methods. From the feature maps obtained from the discriminator, it can be concluded that the ATA-Module contributes to the accuracy improvement because the similarity between pixels is appropriately enhanced by the ATA-Module. 
Therefore, AML-Net with the appropriate combination of Top-Down PDA-Module and ATA-Module can improve the accuracy of IoU for many classes. Deeplabv3+ with ResNet-50 as its backbone does not train well on Drosophila cell images, indicating that even successful models for scene segmentation are not effective for cell image segmentation. In addition, AML-Net has better IoU accuracy than FastFCN using ResNet-50.

The top image group in Figure~\ref{fig:drosophila_Outputs} shows that Top-Down PDA-Module and AML-Net can accurately detect synapses that are easily over-detected by conventional methods. However, our method also caused excessive false positives for cell membrane and mitochondria on the right side. The false positives were probably caused by the fact that the input image shows something very similar to cell membrane and mitochondria. In the lower group of images, U-Net and Top-Down PDA-Module fail to detect mitochondria, and misidentify them as mitochondria in some cell membranes, while ATA-Module and AML-Net correctly identify mitochondria, reducing the number of undetected or false positives. Thus, we can see that AML-Net is able to recognize mitochondria by utilizing the advantages of both ATA-Module and Top-Down PDA-Module.

\begin{figure}
\centering
  \begin{minipage}[b]{.19\linewidth}
    \includegraphics[width=\columnwidth]{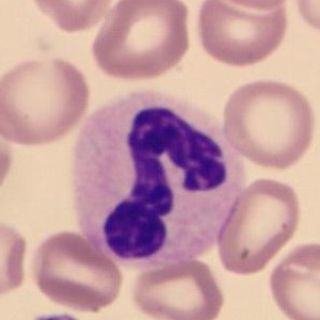}
    \subcaption{}
  \end{minipage}%
  ~
  \begin{minipage}[b]{.19\linewidth}
    \centering
    \includegraphics[width=\columnwidth]{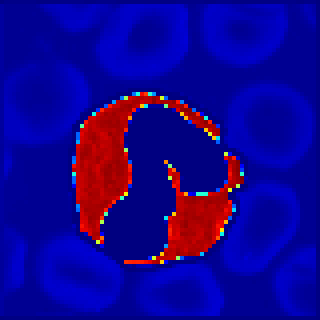}
    \subcaption{}
    \label{b}
  \end{minipage}%
  ~
  \begin{minipage}[b]{.19\linewidth}
    \centering
    \includegraphics[width=\columnwidth]{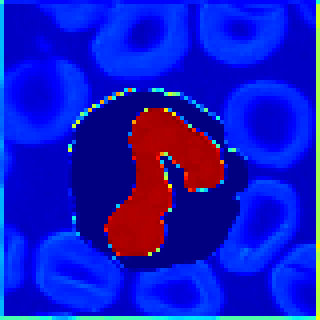}
    \subcaption{}
    \label{c}
  \end{minipage}%
  ~
  \begin{minipage}[b]{.19\linewidth}
    \centering
    \includegraphics[width=\columnwidth]{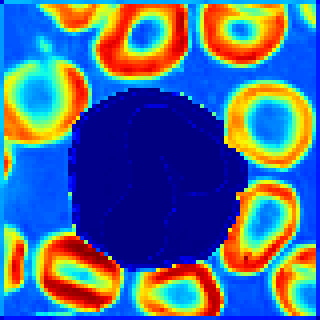}
    \subcaption{}
    \label{d}
  \end{minipage}%
  ~
  \begin{minipage}[b]{.19\linewidth}
    \centering
    \includegraphics[width=\columnwidth]{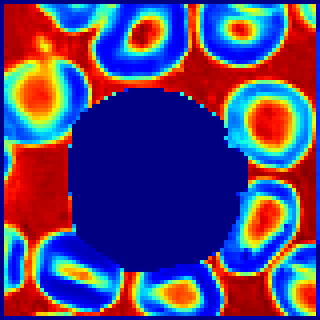}
    \subcaption{}
    \label{e}
  \end{minipage}
  \caption{Visualization results of ATA-Module's Attention Map: (a) Input image. (b-e) Attention map.}
  \label{fig:WBC_Attention}
\end{figure}

\subsubsection{Visualization of ATA-Module}

The attention maps of ATA-Module in AML-Net are visualized in Figure~\ref{fig:WBC_Attention}. When we select a cytoplasmic pixel as a reference, the cytoplasmic regions with high similarity are highlighted in red as shown in Figure~\ref{b}. 
As shown in Figure~\ref{c}, the attention map will respond only to cell nucleus and the other classes are blue with low similarity when cell nucleus is selected as a reference.
In addition, as shown in Figure~\ref{d}, when a pixel of an red blood cell which is a part of the background pixel is selected, only the similar red blood cells are reacted to not the entire background. Therefore, in Figure~\ref{e}, it can be confirmed that red blood cells are not reacted when the pixels of the background other than red blood cells are selected. These results show that ATA-Module can enhance the similarity not only among the three classes; cytoplasm, cell nucleus and background, but also between red blood cells and other parts of the background. Based on these attention maps created from the discriminator, efficient leakage to the generator is performed.

\begin{figure}
\centering
  \begin{minipage}[b]{.19\linewidth}
    \centering
    \includegraphics[width=\columnwidth]{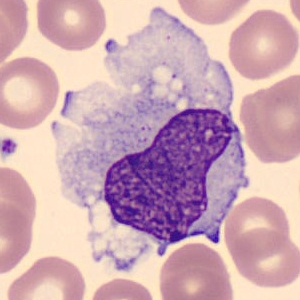}
    \centerline{Input}
  \end{minipage}%
  ~
  \begin{minipage}[b]{.19\linewidth}
    \centering
    \includegraphics[width=\columnwidth]{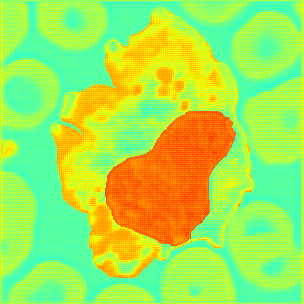}
    \centerline{Epoch=1}
  \end{minipage}%
  ~
  \begin{minipage}[b]{.19\linewidth}
    \centering
    \includegraphics[width=\columnwidth]{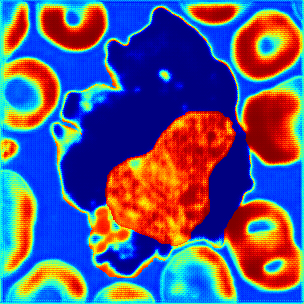}
    \centerline{Epoch=2}
  \end{minipage}%
  ~
  \begin{minipage}[b]{.19\linewidth}
    \centering
    \includegraphics[width=\columnwidth]{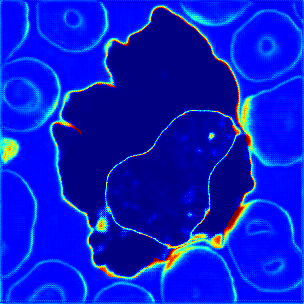}
    \centerline{Epoch=3}
  \end{minipage}%
  ~
  \begin{minipage}[b]{.19\linewidth}
    \centering
    \includegraphics[width=\columnwidth]{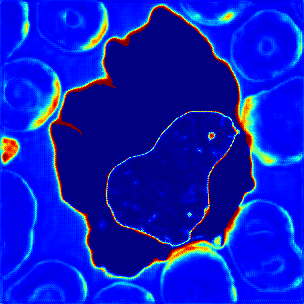}
    \centerline{Epoch=4}
  \end{minipage}
  ~
  \\
  
  \begin{minipage}[b]{.19\linewidth}
    \centering
    \includegraphics[width=\columnwidth]{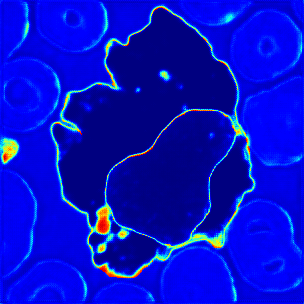}
    \centerline{Epoch=5}
  \end{minipage}%
  ~
  \begin{minipage}[b]{.19\linewidth}
    \centering
    \includegraphics[width=\columnwidth]{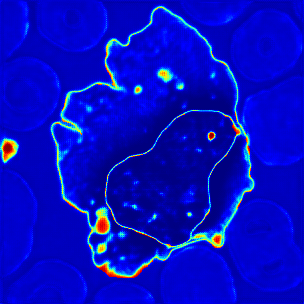}
    \centerline{Epoch=10}
  \end{minipage}%
  ~
  \begin{minipage}[b]{.19\linewidth}
    \centering
    \includegraphics[width=\columnwidth]{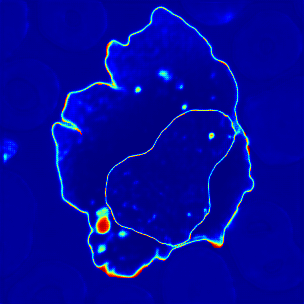}
    \centerline{Epoch=20}
  \end{minipage}%
  ~
  \begin{minipage}[b]{.19\linewidth}
    \centering
    \includegraphics[width=\columnwidth]{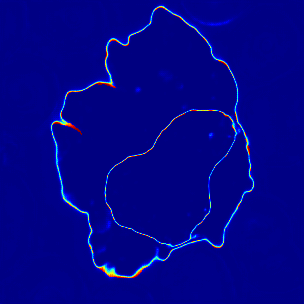}
    \centerline{Epoch=50}
  \end{minipage}%
  ~
  \begin{minipage}[b]{.19\linewidth}
    \centering
    \includegraphics[width=\columnwidth]{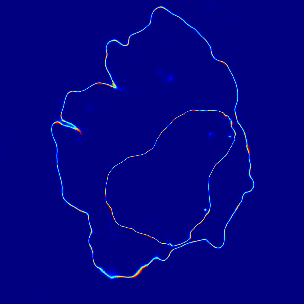}
    \centerline{Epoch=100}
  \end{minipage}
  \caption{Attention map in Top-Down PDA-Module}
  \label{fig:WBC_Probability}
\end{figure}

\subsubsection{Visualization of Top-Down PDA-Module}

Visualization results of the attention map in Top-Down PDA Module
are shown in Figure~\ref{fig:WBC_Probability}. In the first epoch, the probability of the correct class is low at almost pixels. In the second epoch, the probability of the correct class is low for the cell nucleus and red blood cells in the background. As training advances, we can see that the probability of the correct class is low at the edges and in small regions. This indicates that these pixels are more difficult to recognize than the other pixels. Since the pixels with relatively low probability for the correct class contain a lot of information for identifying the object's outline, they are considered to be of relatively high importance during training.

\subsection{Ablation studies}

To show the effectiveness of the proposed method, we perform various ablation studies on WBC dataset~\cite{zheng2018wbc}.

\subsubsection{Ablation studies on ATA-Module}

At first, we compare different connection methods in ATA-Module, which leaks the information from the discriminator to the generator. Comparison methods are as follows. "Add" is the method which adds feature maps in the discriminator to the generator. "$1 \times 1$~Conv" is the method which uses $1 \times 1$ convolution and then add it to the feature map of generator. In addition, we also compare the method using some attention mechanisms; SE block\cite{senet}(SE block), Light Attention\cite{hiramatsu2020semantic} which enhances in spatial and channel by convolution, and Top-Down PDA-Module which is applied to the discrimiator and leaks to the generator (Top-Down PDA-Module). ATA-Module is also compared with a method that creates an Attention Map from the feature maps of the generator and discriminator(Source-Target-Attention).

Table~\ref{table:ATA_Dis2Gen} shows the results while changing the connection in ATA-Module. We see that the leakage from the discriminator to the generator by our ATA-Module achieved the highest accuracy. This indicates that our proposed method is the most effective for enhancing the important information from the discriminator and transmitting it to the generator. Other attention methods enhance the range of the channel or receptive field, so they cannot take into account the detailed relationship between pixels as in ATA-Module. In addition, Sawada's method\cite{sawada2018} using multiple generators cannot be compared fairly because it requires at least two generators. Therefore, we compare our ATA-module with Sawada's method (Concat) under the same condition using two generators and one discriminator.
ATA-Module using Generator2 and Discriminator1 in Table \ref{table:ATA_Dis2Gen} shows better accuracy. The effectiveness of the ATA-Module is also confirmed in the case of multiple generators.

Table~\ref{table:AML_Dis2Gen} shows the results of changing the connection of ATA-Module included in AML-Net. We can see that the accuracy of our AML-Net is the highest. This indicates that the ATA-Module is also highly effective in the connection mechanism from the discriminator to the generator in adversarial mutual leakage. In addition, not only the average IoU accuracy of each class is superior, but also the standard deviation is lower than that of other methods, indicating that the accuracy is more stable and higher than that of conventional methods.

\subsubsection{Ablation stdies on Top-Down PDA-Module}

We evaluate different connection in Top-Down PDA-module. Table~\ref{table:PDA_Gen2Dis} shows that our Top-Down PDA-Module has the best leakage from the generator to the discriminator. Top-Down PDA-Module and PDA-Module have only one Attention Map. Therefore, the information to be transmitted can be compressed based on the difficulty level, and only important pixels can be enhanced. 
In other methods, attention maps are created between channels and pixels to enhance the feature map in the generator before leakage. 
The accuracy of Top-Down PDA-Module and the PDA-Module differs greatly, and pixels that is recognized incorrectly with high confidence by the PDA-Module are judged to be low difficulty by the PDA-Module and remains incorrect. The Top-Down PDA-Module can accurately judge the difficulty as high even if the pixel is recognized incorrectly with a high probability, resulting in higher accuracy.

In addition, the accuracy of changing the Top-Down PDA-Module included in AML-Net is shown in Table \ref{table:AML_Gen2Dis}. "Concat" uses two discriminators because the number of feature maps changes between the first and the second time when the generator leaks to the discriminator. From Table \ref{table:AML_Gen2Dis} we can see that our AML-Net has the highest accuracy. This indicates that our AML-Net can transfer more effectively than other leakage methods in the connection mechanism from the generator to the discriminator in adversarial mutual leakage. In particular, we can confirm that AML-Net outperforms PDA-Module in both the mean and standard deviation of IoU accuracy. This is an advantage of the top-down attention mechanism that uses correct images over the bottom-up attention map.

\begin{table}[t]
    \centering
    \caption{Ablation study in ATA-Module}
    \fontsize{5.5pt}{8pt}\selectfont
    \label{table:ATA_Dis2Gen}
    \begin{tabular}{l|cccc}
    \hline
Method
& Cytoplasm~[\%]                 & Nucleus~[\%]                   & Background~[\%]                & Mean IoU~[\%] \\ \hline \hline
Add
& 67.06±5.78                          & 89.36±1.79                          & 93.48±2.05                          & 83.30±2.84 \\

$1 \times 1$~Conv
& 65.68±7.00                          & {\color{red} \textbf{90.03}}±1.37   & 92.66±2.76                          & 82.79±3.35 \\

SE block\cite{senet}
& 67.08±5.19                          & {\color{green} \textbf{89.46}}±1.81 & 93.31±1.76                          & 83.28±2.75 \\

Light Attention\cite{hiramatsu2020semantic}
& {\color{green} \textbf{68.34}}±6.46 & {\color{blue} \textbf{89.50}}±2.20  & {\color{green} \textbf{93.84}}±1.98 & {\color{blue} \textbf{83.89}}±3.36 \\

Top-Down PDA-Module
& {\color{blue} \textbf{68.50}}±4.90  & 89.10±2.06                          & {\color{blue} \textbf{94.02}}±1.68  & {\color{green} \textbf{83.87}}±2.56 \\

Source-Target-Attention\cite{transformer}
& 67.36±5.87                          & 89.39±1.83                          & 93.55±2.03                          & 83.44±2.90 \\

ATA-Module
& {\color{red} \textbf{69.12}}±7.65   & 89.31±1.86                          & {\color{red} \textbf{94.04}}±2.70   & {\color{red} \textbf{84.16}}±3.82 \\ \hline

\multicolumn{5}{c}{2~Generators, 1~Discriminator} \\ \hline
Concat\cite{sawada2018}
& 68.77±7.19                          & {\color{red} \textbf{90.44}}±1.57   & 93.66±2.40                          & 84.29±3.54 \\

ATA-Module
& {\color{red} \textbf{74.42}}±5.98   & 89.22±1.87                          & {\color{red} \textbf{96.08}}±1.69 & {\color{red} \textbf{86.58}}±2.87 \\ \hline

    \end{tabular}
    \normalsize 
\end{table}

\begin{table}[t]
    \centering
    \caption{Ablation study in ATA-Module included in AML-Net}
    \fontsize{5.5pt}{8pt}\selectfont
    \label{table:AML_Dis2Gen}
    \begin{tabular}{l|cccc}
    \hline
Method
& Cytoplasm~[\%]                 & Nucleus~[\%]                   & Background~[\%]                & Mean IoU~[\%] \\ \hline \hline
Add
& 76.01±7.05                          & 89.54±2.76                          & 96.45±2.30                          & 87.33±3.41                                 \\

$1 \times 1$~Conv
& 72.18±6.75                          & 90.03±1.39                          & 95.09±2.08                          & 85.77±3.14                                 \\

SE block\cite{senet}
& {\color{green} \textbf{76.49}}±7.89 & 89.31±2.65                          & {\color{green} \textbf{96.60}}±2.04 & 87.46±3.85                                 \\

Light Attention\cite{hiramatsu2020semantic}
& {\color{green} \textbf{76.49}}±7.59 & {\color{blue} \textbf{90.28}}±1.75  & 96.36±2.44                          & {\color{green} \textbf{87.71}}±3.57                                 \\

Top-Down PDA-Module
& {\color{blue} \textbf{78.55}}±4.43  & 89.94±2.39                          & {\color{blue} \textbf{97.16}}±1.06 & {\color{blue} \textbf{88.55}}±2.25 \\

Source-Target-Attention
& 74.97±7.56                          & 89.05±3.16                          & 95.99±2.39                          & 86.67±3.97                                 \\

AML-Net(ours)
& {\color{red} \textbf{81.12}}±5.02   & {\color{red} \textbf{90.81}}±1.56   & {\color{red} \textbf{97.59}}±1.26   & {\color{red} \textbf{89.84}}±2.41 \\ \hline

\multicolumn{5}{c}{2~Generators, 1~Discriminator} \\ \hline

Concat
& 74.42±6.71                          & {\color{green} \textbf{90.14}}±1.50 & 95.70±2.08                          & 86.76±3.00                                 \\

AML-Net(ours)
& 73.31±6.08   & 89.73±1.70   & 95.55±1.94   & 86.20±2.83 \\ \hline

    \end{tabular}
    \normalsize 
\end{table}

\begin{table}[t]
    \centering
    \caption{Ablation study in Top-Down PDA-Module}
    \fontsize{5.5pt}{8pt}\selectfont
    \label{table:PDA_Gen2Dis}
    \begin{tabular}{l|cccc}
    \hline
Method
& Cytoplasm~[\%]                     & Nucleus~[\%]                      & Background~[\%]                     & Mean IoU~[\%] \\ \hline \hline
Add
& 70.86±6.91                         & 88.84±2.47                        & 94.99±2.15                          & 84.90±3.38 \\

Concat
& 69.07±4.41                         & 88.63±1.06                        & 94.42±1.45                          & 84.04±1.99 \\

$1 \times 1$~Conv
& 71.72±6.54& 89.51±1.48 & 95.03±2.14                          & 85.42±3.19\\

SE block\cite{senet}
&{\color{green} \textbf{73.34}}±6.80 & {\color{blue} \textbf{89.66}}±1.07  & {\color{green} \textbf{95.50}}±2.39  & {\color{green} \textbf{86.17}}±3.19 \\

Light Attention\cite{hiramatsu2020semantic}
& 70.45±4.82                        & 89.46±1.65                         & 94.75±1.49                          & 84.89±2.40 \\

Self-Attention\cite{transformer}
& 71.35±6.13                        & 88.62±2.27                         & 95.29±1.89 & 85.08±3.13 \\

Source-Target-Attention\cite{transformer}
& 62.84±8.25                        & 88.48±1.62                         & 91.84±3.76                          & 81.05±4.09 \\

PDA-Module\cite{matsuzuki2019PDA}
&{\color{blue} \textbf{76.39}}±6.86  & {\color{green} \textbf{89.64}}±2.82 & {\color{blue} \textbf{96.68}}±1.68   & {\color{blue} \textbf{87.57}}±3.34 \\

Top-Down PDA-Module
& {\color{red} \textbf{77.40}}±7.70  & {\color{red} \textbf{89.84}}±2.40 & {\color{red} \textbf{96.76}}±2.09  & {\color{red} \textbf{88.00}}±3.73 \\ \hline

    \end{tabular}
    \normalsize 
\end{table}

\begin{table}[t]
    \centering
    \caption{Ablation study in Top-Down PDA-Module included in AML-Net}
    \fontsize{5.5pt}{8pt}\selectfont
    \label{table:AML_Gen2Dis}
    \begin{tabular}{l|cccc}
    \hline
Method
& Cytoplasm~[\%]                     & Nucleus~[\%]                       & Background~[\%]                      & Mean IoU~[\%] \\ \hline \hline
Add
& 63.57±7.98                         & 89.18±1.57                         & 91.75±3.66                           & 81.50±4.07 \\

Concat
& 62.08±7.66                         & 87.52±2.86                         & 92.79±3.13                           & 80.80±3.55 \\

$1 \times 1$~Conv
&69.14±6.24                          & 89.82±1.64                         & 94.07±2.13                           & 84.34±3.14\\

SE block\cite{senet}
&{\color{green} \textbf{72.84}}±4.92 & {\color{blue} \textbf{90.14}}±1.66  & {\color{green} \textbf{95.48}}±1.26 & {\color{green} \textbf{86.15}}±2.41 \\

Light Attention\cite{hiramatsu2020semantic}
& 66.74±7.05                         & 89.15±1.78                          & 93.44±2.60                          & 83.11±3.50 \\

Self-Attention\cite{transformer}
& 57.39±8.07                         & 88.26±1.53                          & 89.68±3.48                          & 78.44±3.96 \\

Source-Target-Attention\cite{transformer}
& 59.31±11.06                        & 87.23±3.13                          & 90.57±5.00                          & 79.03±5.77 \\

PDA-Module\cite{matsuzuki2019PDA}
&{\color{blue} \textbf{75.13}}±6.28  & {\color{green} \textbf{90.13}}±2.38 & {\color{blue} \textbf{96.04}}±1.89  & {\color{blue} \textbf{87.10}}±3.02 \\

AML-Net(ours)
& {\color{red} \textbf{81.12}}±5.02  & {\color{red} \textbf{90.81}}±1.56   & {\color{red} \textbf{97.59}}±1.26   & {\color{red} \textbf{89.84}}±2.41 \\ \hline

    \end{tabular}
    \normalsize 
\end{table}

\begin{table}[t]
    \centering
    \caption{Ablation study with some deletions from Top-Down PDA-Module of AML-Net}
    \fontsize{5.5pt}{8pt}\selectfont
    \label{table:AML_w/o}
    \begin{tabular}{l|cccc}
    \hline
Method
& Cytoplasm~[\%]                    & Nucleus~[\%]                       & Background~[\%]                    & Mean IoU~[\%] \\ \hline \hline

w/o 2Gen Top-Down PDA-Module
& {\color{blue} \textbf{78.75}}±5.93 & {\color{red} \textbf{90.94}}±1.11  & {\color{blue} \textbf{96.92}}±1.59 & {\color{blue} \textbf{88.87}}±2.77 \\

w/o 2Dis Top-Down PDA-Module
& {\color{black} {76.66}}±7.36      & {\color{black} {90.78}}±1.12       & {\color{black} {96.20}}±2.45       & {\color{black} {87.88}}±3.46 \\

AML-Net(ours)
& {\color{red} \textbf{81.12}}±5.02 & {\color{blue} \textbf{90.81}}±1.56 & {\color{red} \textbf{97.59}}±1.26  & {\color{red} \textbf{89.84}}±2.41 \\ \hline

    \end{tabular}
    \normalsize 
\end{table}

\begin{table}[t]
    \centering
    \caption{Hyperparameter sensitivity.}
    \fontsize{5.5pt}{8pt}\selectfont
    \label{table:Hyperparameter}
    \begin{tabular}{l|ccccc}
    \hline
Method
& \rotatebox[origin=c]{90}{Val~Mean~IoU~[\%]}                     & \rotatebox[origin=c]{90}{Cytoplasm~[\%]}                     & \rotatebox[origin=c]{90}{Nucleus~[\%]}                      & \rotatebox[origin=c]{90}{Background~[\%]}                     & \rotatebox[origin=c]{90}{Mean IoU~[\%]} \\ \hline \hline
$\lambda_{adv}=0$
& {\color{blue} \textbf{93.98}}±2.05  & 77.42±6.23                         & 90.32±1.46                          & 96.60±2.03                         & 88.11±2.75 \\

$\lambda_{adv}=0.1$
& {\color{green} \textbf{93.92}}±2.05 & {\color{blue} \textbf{78.19}}±5.13 & {\color{green} \textbf{90.62}}±1.85 & {\color{blue} \textbf{96.88}}±1.44 & {\color{green} \textbf{88.56}}±2.46 \\

$\lambda_{adv}=1$
& 93.16±1.91                          & {\color{blue} \textbf{78.19}}±3.79 & {\color{blue} \textbf{90.74}}±1.73  & {\color{blue} \textbf{96.88}}±1.14 & {\color{blue} \textbf{88.60}}±1.71\\

$\lambda_{adv}=0.01$(ours)
& {\color{red} \textbf{94.10}}±1.81   & {\color{red} \textbf{81.12}}±5.02  & {\color{red} \textbf{90.81}}±1.56   & {\color{red} \textbf{97.59}}±1.26  & {\color{red} \textbf{89.84}}±2.41 \\ \hline

    \end{tabular}
    \normalsize 
\end{table}

\subsubsection{w/o Top-Down PDA-Module}
We confirm the effect on accuracy by not using the Attention Map in the generator or discriminator, which is created by the Top-Down PDA-Module of AML-Net and used in the generator and discriminator. Table~\ref{table:AML_w/o} shows the experimental results of the method that does not use the Attention Map of the Top-Down PDA-Module of AML-Net in the generator and the method that removes the leakage to the discriminator. From Table~\ref{table:AML_w/o}, it can be seen that when even a part of the mechanism of the Top-Down PDA-Module is removed, the accuracy decreases in many classes. From these results, we can confirm that the proposed mechanism contributes to the accuracy improvement. In particular, the accuracy is lower when the leakage of the Attention Map from the generator to the discriminator is removed than when the leakage of the Attention Map to the generator is removed, indicating that the leakage from the generator to the discriminator effectively improves the accuracy in the Top-Down PDA-Module.

\subsubsection{Hyperparameter}

We performed an experiment to investigate the effectiveness of discriminator. Table~\ref{table:Hyperparameter} shows the experimental results when the $\lambda_{adv}$ in loss function is changed. In both the validation and test images, a decrease in accuracy can be observed when the hyperparameter $\lambda_{adv}$ is increased. The accuracy degradation is also observed when $\lambda_{adv}=0$ is used in adversarial mutual leakage. In this case, the discriminator is trained but the adversarial loss of the discriminator is not used to train the generator. The result shows that the adversarial loss from the discriminator to the generator is necessary even in the case of adversarial mutual leakage. From this experiment, the hyperparameter in this paper is set to $\lambda_{adv}=0.01$.

{\small
\bibliographystyle{ieee_fullname}
\bibliography{egbib}

\begin{thebibliography}{10}\itemsep=-1pt

\bibitem{segnet}
Vijay Badrinarayanan, Alex Kendall, and Roberto Cipolla.
\newblock Segnet: A deep convolutional encoder-decoder architecture for image
  segmentation.
\newblock {\em IEEE Transactions on Pattern Analysis and Machine Intelligence},
  39(12):2481--2495, 2017.

\bibitem{deeplabv3}
Liang-Chieh Chen, George Papandreou, Florian Schroff, and Hartwig Adam.
\newblock Rethinking atrous convolution for semantic image segmentation.
\newblock {\em arXiv preprint arXiv:1706.05587}, 2017.

\bibitem{deeplabv3plus}
Liang-Chieh Chen, Yukun Zhu, George Papandreou, Florian Schroff, and Hartwig
  Adam.
\newblock Encoder-decoder with atrous separable convolution for semantic image
  segmentation.
\newblock In {\em Proceedings of the European Conference on Computer Vision},
  pages 801--818, 2018.

\bibitem{choi2019self}
Jaehoon Choi, Taekyung Kim, and Changick Kim.
\newblock Self-ensembling with gan-based data augmentation for domain
  adaptation in semantic segmentation.
\newblock In {\em Proceedings of the IEEE/CVF International Conference on
  Computer VisionProceedings of the IEEE/CVF International Conference on
  Computer VisionProceedings of the IEEE/CVF International Conference on
  Computer Vision}, pages 6830--6840, 2019.

\bibitem{fu2019dual}
Jun Fu, Jing Liu, Haijie Tian, Yong Li, Yongjun Bao, Zhiwei Fang, and Hanqing
  Lu.
\newblock Dual attention network for scene segmentation.
\newblock In {\em Proceedings of the IEEE/CVF Conference on Computer Vision and
  Pattern Recognition}, pages 3146--3154, 2019.

\bibitem{sstem}
Stephan Gerhard, Jan Funke, Julien Martel, Albert Cardona, and Richard Fetter.
\newblock {Segmented Anisotropic ssTEM Dataset of Neural Tissue}.
\newblock {\em
  \url{https://figshare.com/articles/Segmented_anisotropic_ssTEM_dataset_of_neural_tissue/856713}}.

\bibitem{gan}
Ian Goodfellow, Jean Pouget-Abadie, Mehdi Mirza, Bing Xu, David Warde-Farley,
  Sherjil Ozair, Aaron Courville, and Yoshua Bengio.
\newblock Generative adversarial nets.
\newblock In {\em Advances in Neural Information Processing Systems}, pages
  2672--2680, 2014.

\bibitem{resnet}
Kaiming He, Xiangyu Zhang, Shaoqing Ren, and Jian Sun.
\newblock Deep residual learning for image recognition.
\newblock In {\em Proceedings of the IEEE Conference on Computer Vision and
  Pattern Recognition}, pages 770--778, 2016.

\bibitem{hiramatsu2020semantic}
Yuki. Hiramatsu and Kazuhiro Hotta.
\newblock Semantic segmentation using light attention mechanism.
\newblock In {\em Proceedings of the 15th International Joint Conference on
  Computer Vision, Imaging and Computer Graphics Theory and Applications},
  pages 622--625, 2020.

\bibitem{hoffman2017cycada}
Judy Hoffman, Eric Tzeng, Taesung Park, Jun-Yan Zhu, Phillip Isola, Kate
  Saenko, Alexei~A Efros, and Trevor Darrell.
\newblock Cycada: Cycle-consistent adversarial domain adaptation.
\newblock In {\em International Conference on Machine Learning}, pages
  1994--2003, 2017.

\bibitem{hoover2000STARE}
AD Hoover, Valentina Kouznetsova, and Michael Goldbaum.
\newblock Locating blood vessels in retinal images by piecewise threshold
  probing of a matched filter response.
\newblock {\em IEEE Transactions on Medical imaging}, 19(3):203--210, 2000.

\bibitem{senet}
Jie Hu, Li Shen, and Gang Sun.
\newblock Squeeze-and-excitation networks.
\newblock In {\em Proceedings of the IEEE Conference on Computer Vision and
  Pattern Recognition}, pages 7132--7141, 2018.

\bibitem{huang2019ccnet}
Zilong Huang, Xinggang Wang, Lichao Huang, Chang Huang, Yunchao Wei, and Wenyu
  Liu.
\newblock Ccnet: Criss-cross attention for semantic segmentation.
\newblock In {\em Proceedings of the IEEE/CVF International Conference on
  Computer VisionProceedings of the IEEE/CVF International Conference on
  Computer VisionProceedings of the IEEE/CVF International Conference on
  Computer Vision}, pages 603--612, 2019.

\bibitem{imanishi2018novel}
Ayako Imanishi, Tomokazu Murata, Masaya Sato, Kazuhiro Hotta, Itaru Imayoshi,
  Michiyuki Matsuda, and Kenta Terai.
\newblock A novel morphological marker for the analysis of molecular activities
  at the single-cell level.
\newblock {\em Cell Structure and function}, pages 129--140, 2018.

\bibitem{pix2pix}
Phillip Isola, Jun-Yan Zhu, Tinghui Zhou, and Alexei~A Efros.
\newblock Image-to-image translation with conditional adversarial networks.
\newblock In {\em Proceedings of the IEEE Conference on Computer Vision and
  Pattern Recognition}, pages 1125--1134, 2017.

\bibitem{liao2019evaluate}
Fangzhou Liao, Ming Liang, Zhe Li, Xiaolin Hu, and Sen Song.
\newblock Evaluate the malignancy of pulmonary nodules using the 3-d deep leaky
  noisy-or network.
\newblock {\em IEEE transactions on neural networks and learning systems},
  30(11):3484--3495, 2019.

\bibitem{lin2017refinenet}
Guosheng Lin, Anton Milan, Chunhua Shen, and Ian Reid.
\newblock Refinenet: Multi-path refinement networks for high-resolution
  semantic segmentation.
\newblock In {\em Proceedings of the IEEE Conference on Computer Vision and
  Pattern Recognition}, pages 1925--1934, 2017.

\bibitem{liu2016coupled}
Ming-Yu Liu and Oncel Tuzel.
\newblock Coupled generative adversarial networks.
\newblock In {\em Advances in Neural Information Processing Systems}, pages
  469--477, 2016.

\bibitem{fcn}
Jonathan Long, Evan Shelhamer, and Trevor Darrell.
\newblock Fully convolutional networks for semantic segmentation.
\newblock In {\em Proceedings of the IEEE Conference on Computer Vision and
  Pattern Recognition}, pages 3431--3440, 2015.

\bibitem{luc2016semantic}
Pauline Luc, Camille Couprie, Soumith Chintala, and Jakob Verbeek.
\newblock Semantic segmentation using adversarial networks.
\newblock In {\em Advances in Neural Information Processing Systems Workshops},
  2016.

\bibitem{matsuzuki2019PDA}
Daisuke Matsuzuki and Kazuhiro Hotta.
\newblock Cell image segmentation using attention module at each layer.
\newblock In {\em Proceedings of The 20th International Conference on Systems
  Biology}, 2019.

\bibitem{miyato2018spectral}
Takeru Miyato, Toshiki Kataoka, Masanori Koyama, and Yuichi Yoshida.
\newblock Spectral normalization for generative adversarial networks.
\newblock In {\em In International Conference on Learning Representations},
  2018.

\bibitem{oktay2018attention_unet}
Ozan Oktay, Jo Schlemper, Loic~Le Folgoc, Matthew Lee, Mattias Heinrich,
  Kazunari Misawa, Kensaku Mori, Steven McDonagh, Nils~Y Hammerla, Bernhard
  Kainz, et~al.
\newblock Attention u-net: Learning where to look for the pancreas.
\newblock In {\em Medical Imaging with Deep Learning}, 2018.

\bibitem{owen2009CHASE}
Christopher~G Owen, Alicja~R Rudnicka, Robert Mullen, Sarah~A Barman, Dorothy
  Monekosso, Peter~H Whincup, Jeffrey Ng, and Carl Paterson.
\newblock Measuring retinal vessel tortuosity in 10-year-old children:
  Validation of the computer-assisted image analysis of the retina (caiar)
  program.
\newblock {\em Investigative ophthalmology \& visual science},
  50(5):2004--2010, 2009.

\bibitem{park2019semantic}
Taesung Park, Ming-Yu Liu, Ting-Chun Wang, and Jun-Yan Zhu.
\newblock Semantic image synthesis with spatially-adaptive normalization.
\newblock In {\em Proceedings of the IEEE/CVF Conference on Computer Vision and
  Pattern Recognition}, pages 2337--2346, 2019.

\bibitem{stand-alone}
Prajit Ramachandran, Niki Parmar, Ashish Vaswani, Irwan Bello, Anselm Levskaya,
  and Jonathon Shlens.
\newblock Stand-alone self-attention in vision models.
\newblock In {\em Advances in Neural Information Processing Systems}, pages
  68--80, 2019.

\bibitem{reed2016generative}
Scott Reed, Zeynep Akata, Xinchen Yan, Lajanugen Logeswaran, Bernt Schiele, and
  Honglak Lee.
\newblock Generative adversarial text to image synthesis.
\newblock In {\em International Conference on Machine Learning}, 2016.

\bibitem{unet}
Olaf Ronneberger, Philipp Fischer, and Thomas Brox.
\newblock U-net: Convolutional networks for biomedical image segmentation.
\newblock In {\em International Conference on Medical Image Computing and
  Computer-Assisted Intervention}, pages 234--241, 2015.

\bibitem{roy2018scsenet}
Abhijit~Guha Roy, Nassir Navab, and Christian Wachinger.
\newblock Concurrent spatial and channel ‘squeeze \& excitation’in fully
  convolutional networks.
\newblock In {\em International Conference on Medical Image Computing and
  Computer-Assisted Intervention}, pages 421--429, 2018.

\bibitem{saito2018maximum}
Kuniaki Saito, Kohei Watanabe, Yoshitaka Ushiku, and Tatsuya Harada.
\newblock Maximum classifier discrepancy for unsupervised domain adaptation.
\newblock In {\em Proceedings of the IEEE Conference on Computer Vision and
  Pattern Recognition}, pages 3723--3732, 2018.

\bibitem{sawada2018}
Kyoya Sawada, Kazuhiro Hotta, Ayako Imanishi, Michiyuki Matsuda, and Kenta
  Terai.
\newblock Segmentation of cell images by leaking the information of
  discriminator to generator.
\newblock In {\em International Conference of the IEEE Engineering in Medicine
  and Biology Society}, 2018.

\bibitem{accuracy-booster}
Pravendra Singh, PRATIK MAZUMDER, and Vinay Namboodiri.
\newblock Accuracy booster: Performance boosting using feature map
  re-calibration.
\newblock In {\em Proceedings of the IEEE/CVF Winter Conference on Applications
  of Computer Vision}, pages 884--893, 2020.

\bibitem{2004DRIVE}
Joes Staal, Michael~D Abr{\`a}moff, Meindert Niemeijer, Max~A Viergever, and
  Bram Van~Ginneken.
\newblock Ridge-based vessel segmentation in color images of the retina.
\newblock {\em IEEE transactions on medical imaging}, 23(4):501--509, 2004.

\bibitem{taigman2017unsupervised}
Yaniv Taigman, Adam Polyak, and Lior Wolf.
\newblock Unsupervised cross-domain image generation.
\newblock In {\em In International Conference on Learning Representations},
  2017.

\bibitem{tzeng2017adversarial}
Eric Tzeng, Judy Hoffman, Kate Saenko, and Trevor Darrell.
\newblock Adversarial discriminative domain adaptation.
\newblock In {\em Proceedings of the IEEE Conference on Computer Vision and
  Pattern Recognition}, pages 7167--7176, 2017.

\bibitem{transformer}
Ashish Vaswani, Noam Shazeer, Niki Parmar, Jakob Uszkoreit, Llion Jones,
  Aidan~N Gomez, {\L}ukasz Kaiser, and Illia Polosukhin.
\newblock Attention is all you need.
\newblock In {\em Advances in Neural Information Processing Systems}, pages
  5998--6008, 2017.

\bibitem{wang2017residual}
Fei Wang, Mengqing Jiang, Chen Qian, Shuo Yang, Cheng Li, Honggang Zhang,
  Xiaogang Wang, and Xiaoou Tang.
\newblock Residual attention network for image classification.
\newblock In {\em Proceedings of the IEEE Conference on Computer Vision and
  Pattern Recognition}, pages 3156--3164, 2017.

\bibitem{wang2019eca}
Qilong Wang, Banggu Wu, Pengfei Zhu, Peihua Li, Wangmeng Zuo, and Qinghua Hu.
\newblock Eca-net: Efficient channel attention for deep convolutional neural
  networks.
\newblock {\em arXiv preprint arXiv:1910.03151}, 2019.

\bibitem{wang2018non_local}
Xiaolong Wang, Ross Girshick, Abhinav Gupta, and Kaiming He.
\newblock Non-local neural networks.
\newblock In {\em Proceedings of the IEEE Conference on Computer Vision and
  Pattern Recognition}, pages 7794--7803, 2018.

\bibitem{wu2019fastfcn}
Huikai Wu, Junge Zhang, Kaiqi Huang, Kongming Liang, and Yu Yizhou.
\newblock Fastfcn: Rethinking dilated convolution in the backbone for semantic
  segmentation.
\newblock In {\em arXiv preprint arXiv:1903.11816}, 2019.

\bibitem{xue2018segan}
Yuan Xue, Tao Xu, Han Zhang, L~Rodney Long, and Xiaolei Huang.
\newblock Segan: Adversarial network with multi-scale l1 loss for medical image
  segmentation.
\newblock {\em Neuroinformatics}, 16(3-4):383--392, 2018.

\bibitem{dilatedconv}
Fisher Yu and Vladlen Koltun.
\newblock Multi-scale context aggregation by dilated convolutions.
\newblock In {\em In International Conference on Learning Representations},
  2016.

\bibitem{sagan}
Han Zhang, Ian Goodfellow, Dimitris Metaxas, and Augustus Odena.
\newblock Self-attention generative adversarial networks.
\newblock In {\em International Conference on Machine Learning}, pages
  7354--7363, 2019.

\bibitem{zhang2017stackgan}
Han Zhang, Tao Xu, Hongsheng Li, Shaoting Zhang, Xiaogang Wang, Xiaolei Huang,
  and Dimitris~N Metaxas.
\newblock Stackgan: Text to photo-realistic image synthesis with stacked
  generative adversarial networks.
\newblock In {\em Proceedings of the IEEE International Conference on Computer
  Vision}, pages 5907--5915, 2017.

\bibitem{zheng2018wbc}
Xin Zheng, Yong Wang, Guoyou Wang, and Jianguo Liu.
\newblock Fast and robust segmentation of white blood cell images by
  self-supervised learning.
\newblock {\em Micron}, 107:55--71, 2018.

\bibitem{cycle}
Jun-Yan Zhu, Taesung Park, Phillip Isola, and Alexei~A Efros.
\newblock Unpaired image-to-image translation using cycle-consistent
  adversarial networks.
\newblock In {\em Proceedings of the IEEE International Conference on Computer
  Vision}, pages 2223--2232, 2017.

\end{thebibliography}
}

\end{document}